%% file: QuaStar.tex
\documentclass[iop,twocolumn]{emulateapj}

\usepackage{natbib,amsmath}
\usepackage{graphicx}
\usepackage{natbib}
\usepackage{amssymb}
\usepackage{float}
\usepackage{makeidx}
\usepackage{amsmath}
\usepackage{rotating}
\usepackage{longtable}
\usepackage{subfigure}
\usepackage{morefloats}
\usepackage{captcont}
\usepackage{xcolor} 
\usepackage{chngcntr}
\usepackage{wrapfig}
\usepackage{multirow}

\usepackage{wasysym}
\newcommand{\QSOexcessdot}[1][6]{{\fontsize{#1}{0}\selectfont\raisebox{1pt}{\CIRCLE}}}
\newcommand{\BHBexcessdot}[1][12]{{\fontsize{#1}{0}\selectfont\raisebox{-0.5pt}{$\circ$}}}
 
\graphicspath { {./}}

\newcommand{\etal}{et~al.\/}

\newcommand{\HI}{{\sc Hi}\,}

\newcommand{\kms}{\mbox{km\thinspace s$^{-1}$\,}}

\newcommand{\Mstar}{\mbox{${\cal M}_\star$}}

\newcommand{\percmsq}{cm$^{-2}$ }

\newcommand{\OVI}{\hbox{O\,{\sc vi}}\,}

\newcommand{\SiII}{\hbox{Si\,{\sc ii}}\,}
\newcommand{\SiIII}{\hbox{Si\,{\sc iii}}\,}
\newcommand{\SiIV}{\hbox{Si\,{\sc iv}}\,}

\newcommand{\CII}{\hbox{C\,{\sc ii}}\,}
\newcommand{\CIV}{\hbox{C\,{\sc iv}}\,}
\newcommand{\NV}{\hbox{N\,{\sc v}}\,}
\newcommand{\NaI}{\hbox{Na\,{\sc i}}\,}
\newcommand{\CaII}{\hbox{Ca\,{\sc ii}}\,}
\newcommand\mlstar{L^*}
\newcommand\lstar{$\mlstar$}

\newcommand{\diffLVCGM}{$\rm \Delta  logN_{LVCGM}$\,}

\shortauthors{Bish \etal}
\shorttitle{Low-Velocity Gas in the Galactic CGM}

\begin{document}
\title{The QuaStar Survey: Detecting Hidden Low-Velocity Gas in the Milky Way's Circumgalactic Medium}

\author{Hannah V. Bish\altaffilmark{1}, 
Jessica K. Werk\altaffilmark{1}, 
Joshua Peek\altaffilmark{2,3}, 
Yong Zheng\altaffilmark{4,5}, 
Mary Putman\altaffilmark{6} 
}

\altaffiltext{1}{Department of Astronomy, University of Washington, Seattle, WA 98195, USA; $hvbish@uw.edu$}
\altaffiltext{2}{Space Telescope Science Institute, 3700 San Martin Drive, Baltimore, MD 21218, USA}
\altaffiltext{3}{Department of Physics \& Astronomy, 
Johns Hopkins University,
3400 N. Charles Street, 
Baltimore, MD 21218}
\altaffiltext{4}{Department of Astronomy, University of California, Berkeley, CA 94720, USA}
\altaffiltext{5}{Miller Institute for Basic Research in Science, University of California,  Berkeley, CA 94720, USA}
\altaffiltext{6}{Department of Astronomy, Columbia University, New York, NY 10027, USA}

\begin{abstract}

From our position embedded within the Milky Way's interstellar medium (ISM), we have limited ability to detect gas at low relative velocities in the extended Galactic halo because those spectral lines are blended with much stronger signals from dense foreground gas. As a result, the content of the Milky Way's circumgalactic medium (CGM) is poorly constrained at $|v_{\rm LSR}|$ $\lesssim$ 150 km s$^{-1}$. To overcome this complication, the QuaStar Survey applies a spectral differencing technique using paired quasar-star sightlines to measure the obscured content of the Milky Way's CGM for the first time. We present measurements of the \CIV doublet ($\lambda\lambda$ 1548, 1550 \AA), a rest-frame UV metal line transition detected in HST/COS spectra of 30 halo-star/quasar pairs evenly distributed across the sky at Galactic latitudes $|b|>30^\circ$. The 30 halo stars have well-constrained distances (d$\approx$5-14 kpc), and are paired with quasars separated by $<$ 2.8$^\circ$. We argue that the difference in absorption between the quasar and stellar sightlines originates primarily in the Milky Way's extended CGM beyond $\sim$10 kpc. For the Milky Way's extended, low-velocity CGM ($|v|<$150 km/s), we place an upper limit on the mean \CIV column density of \diffLVCGM$< 13.39$ and find a covering fraction of $f_{\rm CIV,LVCGM} (\rm logN>13.65)=$ 20\% [6/30], a value significantly lower than the covering fraction for star-forming galaxies at low redshift. Our results suggest either that the bulk of Milky Way's \CIV-traced CGM lies at low Galactic latitudes, or that the Milky Way's CGM is lacking in warm, ionized material compared to low-redshift ($z < 0.1$) star-forming galaxy halos.\\

\end{abstract}


\section{Introduction}
\label{sec:intro}

Over the past decade, groundbreaking work with the Cosmic Origins Spectrograph (COS) on the Hubble Space Telescope (HST) has shown that the circumgalactic medium (CGM) around Milky Way-like galaxies is the hiding place of most galactic baryons and metals and an important source of fuel for star formation \citep{werk2014,peeples2014,tumlinsonARAA2017}. This tenuous, predominantly ionized halo gas is notoriously difficult to detect in emission because of its low density, so it is most commonly measured in absorption along quasar sightlines that pierce nearby galaxy halos \citep[e.g.][]{bahcallspitzer1969, bergeron1986, lanzetta1995, prochaska2011, werk2013}. Extragalactic absorption-line studies are useful for investigating average global properties of the CGM, but the limited availability of quasar sightlines means that we can rarely make more than one or two measurements within a single galaxy halo outside the Local Group. The Milky Way, with hundreds of sightlines that pierce its halo, provides a unique opportunity to characterize the structure and kinematics of our own CGM better than that of any other galaxy in the universe.

Observations of CGM gas reveal a complex structure and distribution that can be explained by various physical models. On the largest scales, the CGM may be distributed spherically with a steeply declining density profile (e.g. \citealt{liangchen2014, werk2014}); it may form a thick, rotating, toroidal structure (e.g. \citealt{Stewart2011,bregman2018}); or it may be composed of large shocked sheets of material flowing through the halo \citep{mcquinnwerk2018}. The northern and southern Galactic skies are quite different in their known CGM components, which may indicate an overall asymmetry \citep{putman2012,bordoloi2017}. Additionally, smaller-scale structures, such as clouds and thin filaments, are known to populate the CGM \citep[e.g.][]{bekhti2009, saul2012,stocke2013, mccourt2015, stern2016, bish2019}.

While our own Galaxy was the first in which we detected extraplanar material \citep{muller1963}, we have yet to make a definitive measurement of the column density of the Milky Way’s CGM because global measurements of Galactic halo properties are confounded by our location within the Milky Way’s disk. Any attempt to isolate signals from the Milky Way's CGM will be hampered by blended foreground absorption from the interstellar medium (ISM) in which we are embedded (see \citealt{zheng2015}). In particular, Galactic foreground gas in the disk obscures halo gas with similar relative velocities.

In order to avoid foreground contamination from the low-velocity material in the disk, Milky Way CGM studies typically target high-velocity gas ($|v|\gtrsim 100$ km/s) and do not constrain lower-velocity material beyond the inner $\sim$10 kpc of the halo. Ionized high-velocity clouds are ubiquitous, covering 65\%-90\% of the sky at typical distances of 5-15 kpc \citep{fox2006, collins2009, shull2009, lehner2012, richter2017}. But despite their abundance, observations indicate that high- and intermediate-velocity clouds likely make up a small fraction of the Milky Way's total halo mass \citep{putman2012} and point to the possibility that considerable material is hidden at lower velocities in the extended Galactic CGM.

In other low-redshift galaxies, the average velocity of galactic CGM absorption generally centers around 0 km/s relative to the host galaxy's systemic velocity \citep[e.g.][]{tumlinson2013, werk2016}. Metal-enriched gas is detected in external \lstar{} galaxies moving at low relative velocities out to their virial radii \citep[e.g.,][]{liangchen2014,werk2014,lehner2015,Burchett2016}. In the COS-Halos survey of low-redshift galaxy halos, $\sim$ 90\% of the total CGM column density within $\pm600$ km/s of the galaxy systemic velocity was found at  $|v| < 100$ km/s \citep{tumlinson2013,werk2014}. Kinematic analysis of warm gas signatures in the quasar absorption-line survey COS-GAL also shows that a substantial portion of the CGM is likely hidden at $|v| < 100$ km/s \citep{zheng2019}. In the recent CGM$^2$ survey, the majority of the \HI~ column density detected in absorption around $z<0.48$ galaxies lies within $\pm250$ km s$^{-1}$ of the galaxy systemic velocity, despite a velocity window of 1000 km s$^{-1}$ \citep{wilde2021}. Additionally, \cite{martinho2019} find that the velocity of low-ionization gas around nearby galaxies is tied to the rotation of the disk out to radii of $\sim$70 kpc. If the Milky Way is similar to other star-forming galaxies at low redshift, these observations further reinforce the evidence suggesting that a significant fraction of Galactic baryons are yet to be detected in the low-velocity Galactic CGM.

Simulations of Milky Way analogs also broadly agree that a significant fraction of circumgalactic baryons are moving at low velocities. The FOGGIE simulations, which focus on resolving the CGM around galaxies with extreme spatial and mass resolution, show with mock observations that the strongest absorption is confined to within 200 km/s of the systemic velocity \citep{peeples2019, zheng2020}.
Other synthetic observations of a Milky Way-mass galaxy generated from the high-resolution cosmological simulation Enzo also predict that $\sim65\%$ of the Milky Way’s CGM mass at Galactic latitudes $|b|$ $>$ 20$^{\circ}$ is hidden at low to intermediate velocities ($|v| <$ 100 km/s) for all gas phases and that the warm-hot component ($10^5$ K $< T <$ $10^6$ K) is primarily moving at $|v| <$ 150 km/s \citep{zheng2015, joung2012}.

The \CIV doublet is a useful probe of warm ionized gas in galaxy halos because of its strong oscillator strength \citep{savage2009} and high covering fraction in the CGM of external galaxies \citep{keeney2013, liangchen2014}. In low-redshift galaxies that span a wide range of stellar mass (9.5 $<$ log ($M_*/M_\odot$) $<$ 11), estimates of log $\rm N_{CIV} ~[cm^{-2}]$ range from 13.5 - 14.5, with a covering fraction of $\sim$50\% out to half the virial radius for log$\rm N_{CIV}$ $>$ 13.5 \percmsq \citep{Bordoloi2014,Burchett2016}. Using conservative ionization fraction arguments which take the maximum fraction of carbon that presents as \CIV, $\sim$30\%, for both photoionized and collisionally ionized gas, such measurements imply $> 10^6 M_{\odot}$ of carbon sits in the CGM of external galaxies. For sub-\lstar{} galaxies with stellar masses 10$^{8.5}$ $\lesssim$ \Mstar{} [M$_{\odot}$] $\lesssim$ 10$^{10}$, this estimated CGM carbon mass amounts to 50\%-80\% of the total ISM carbon mass \citep{Bordoloi2014}.

Measurements of the Milky Way's CGM are needed at all velocities in order to directly probe the baryonic mass of the Milky Way, test the expectations from cosmological simulations, and compare the Milky Way to other galaxies. Motivated by this need, we present QuaStar: an absorption-line survey designed to enable the first measurements of low-velocity CGM gas in the Milky Way that have been isolated from contaminating foreground absorption.

The QuaStar survey follows recent work examining ionized CGM gas absorption in the Milky Way using HST/COS archival spectra of hundreds of quasars \citep{richter2017, zheng2019, qu2020}. \cite{richter2017} provide a list of quasar pairs with $<1^\circ$ sightline separation and highlight the differences in their \SiII, \SiIII, \CII, and \CIV absorption profiles. Our preliminary investigation of sightlines included in the \cite{zheng2019} COS-GAL sample showed that the variation of \CIV for 20/26 quasar pairs with angular separation $<2^\circ$ is consistent with noise. With this study we build on the discovery that \CIV, sensitive to warm (T $>$ 10$^{4.5}$ K) CGM gas, is found in large coherent structures and exhibits little variation in column density between sightlines at small separation \citep{werk2019}.

Specifically,  QuaStar uses close pairs of quasar and halo star sightlines to account for the foreground gas absorption that blends with and obscures our view of low-velocity gas in the Milky Way's CGM. The experimental design of QuaStar relies on the expectation that \CIV-bearing gas in the inner halo of the Milky Way is homogeneous on the angular scales probed by the star-quasar sightline pairs -- an assumption that we seek to evaluate as we carry out the analysis. If both quasar and stellar spectra have the same amount of \CIV absorption due to foreground gas, then the difference in total \CIV absorption between them likely originates in the Milky Way's extended CGM, revealing its unobscured content for the first time.

Here we focus on the first isolated measurement of N$_{\rm CIV}$ in the extended Milky Way halo (R$_{\rm CGM}$ $\gtrsim$ 10 kpc) for 30 sightline pairs across the sky and use our findings to place the Milky Way's CGM content in a broader extragalactic context. \S\ref{sec:observations} expands on the details of the QuaStar survey design, sample, and data. \S\ref{sec:analysis} describes our analysis methods for obtaining column density measurements and applying a ``spectral differencing" technique to isolate low-velocity \CIV in the Milky Way's CGM. We present the results of our ISM-subtracted \CIV measurements in \S\ref{sec:results} and discuss the implications of our findings in \S\ref{sec:discussion}. Finally, we conclude with a summary in \S\ref{sec:summary}. \\
 \\

\section{Observations}
\label{sec:observations}

\subsection{Survey Design}

The QuaStar survey consists of HST/COS G160M spectroscopy of 30 UV-bright halo stars at distances $d\sim5-14$ kpc paired with archival G160M spectra of quasars separated by less than $2.8^\circ$ on the sky (Figure \ref{fig:experimentaldesign} and Table \ref{table:properties}).
Quasar sightlines probe the extended CGM, while halo star sightlines span foreground absorption from the disk and the disk-halo interface. Thus, the survey design allows absorption along the stellar sightline (originating from foreground gas associated with the disk) to be subtracted from absorption along the quasar sightline (originating from both disk and CGM gas), in principle isolating absorption signatures of gas beyond the disk-halo interface.\\

\subsection{Sample Selection}

To ensure that foreground ISM contamination could be subtracted successfully, we selected a sample of halo stars that are (1) far-UV (FUV) bright (FUV magnitude $<$ 17.5), (2) well above the Galactic plane ($|z|$ $>$ 6 kpc) and (3) close on the sky ($<2.8^\circ$) to quasars with well-detected rest-frame \CIV. The disk-halo interface has been suggested to have a scale height of $\sim$3 kpc in \CIV \citep{savage2009}, so we targeted stars $>$6 kpc above the Galactic plane. Assuming an exponential density profile along the $z$-axis for \CIV-bearing gas, a halo star at $z\sim$6 kpc allows us to subtract out $\gtrsim$86\% ($= 1 - e^{6/3}$) of absorption from the obscuring disk-associated material.

Although sightlines at lower Galactic latitudes would be sensitive to CGM features in the extended plane of the disk, UV-bright quasar lines of sight at low Galactic latitudes are rare due to significant extinction from the Milky Way disk. Generally, sightlines at higher Galactic latitudes are less affected by Galactic differential rotation and generally must pass through less foreground ISM substructure in order to probe gas at $>$6 kpc above the disk. As a result, we chose QuaStar lines of sight to lie at $|b|$ $>$ 30$^{\circ}$. We sampled the sightlines as evenly as possible in Galactic longitude and within our range of Galactic latitudes. In selecting lines of sight, we did not avoid the well-studied, ionized Magellanic System \citep[MS;][]{fox2014}, the bulk of which lies at d $\gtrsim$ 50 kpc and $|v_{\rm LSR}|$ $>$ 100 km s$^{-1}$. We discuss its possible contribution to our \CIV measurements throughout the manuscript, and sightlines that intersect \HI associated with the MS are labeled ``MS" in Figure \ref{fig:difference-spectra} and Table \ref{table:properties}.

\begin{figure}[h]
    \centering
    \vspace{0.7cm}
    \includegraphics[width=0.44\textwidth]{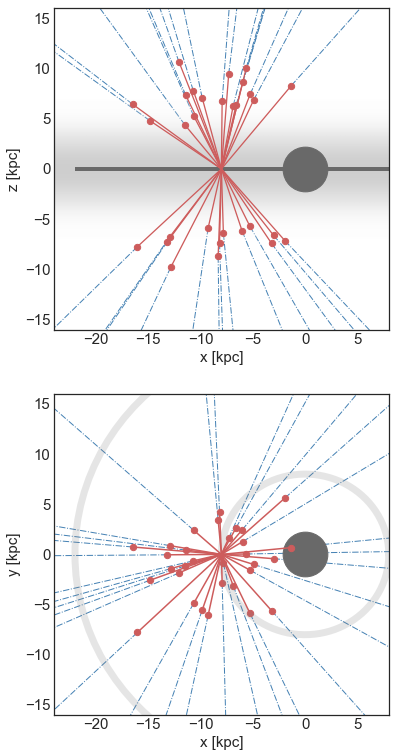}
    \vspace{0.2cm}
    \caption{{\sc QuaStar survey design.} Spatial distribution of close star-quasar sightline pairs  ($<2.8^{\circ}$ separation) in the QuaStar survey, projected onto the $x$-$z$ (top) and $x$-$y$ (bottom) planes of the Milky Way disk. Sightlines toward halo stars are shown in red, and the blue dashed lines show the corresponding sightlines to the paired quasar. The Galactic disk and bulge are shown in dark gray for reference. In the top panel, the approximate extent of the disk-halo interface is shaded light gray \citep{savage2009}. In the bottom panel, lines of constant Galactocentric radius are shown in gray.\\} 
    \label{fig:experimentaldesign}
\end{figure}

The stellar sightlines for QuaStar are blue horizontal branch halo (BHB) stars, which have little scatter in their absolute magnitudes and generally exhibit FUV-bright continua with few stellar absorption features \citep{deason2011, werk2019}.  To construct a parent sample of halo stars, we started with the \cite{smith2010} catalog of BHB candidates. This catalog was itself constructed from photometric Sloan Digital Sky Survey (SDSS) observations, using a machine learning method trained on the spectroscopically confirmed catalog of \cite{xue2008}. We then matched the stars to the Gaia DR2 catalog for parallaxes \citep{gaiaDR2mission,gaiaDR2summary}. While Gaia does not have the parallactic sensitivity to robustly measure the distance to these halo stars, the primary interlopers into our color selection are nearby disk stars, which Gaia parallaxes allow us to reject. We found that these interloper disk stars composed roughly 20\% of the original sample. 

The SDSS sample is largely limited to the northern sky, so for southern targets we turned to the SkyMapper survey \citep{skymapper-DR1}. SkyMapper is the only large-area southern survey with accurate $u$-band photometry, a critical filter for selecting BHB stars. We found a small subset of SDSS BHB stars from \cite{smith2010} that are also detected in SkyMapper. These served as the training set for a k-nearest neighbors algorithm, which we used to predict the BHB probability of the rest of the SkyMapper stellar sample based on colors. Using the same Gaia method to reject interlopers, we found the highest likelihood BHB ($>$90\%) subsample to be 10\% disk stars, which we rejected.

We then took this decontaminated joint SDSS/SkyMapper parent sample of halo BHB stars and found all stars with well-measured FUV magnitudes from GALEX \citep{galex}. We took the subset of these stars brighter than FUV = 18.5 and cross-matched it with the COS-GAL sample of quasars \citep{zheng2019} with well-measured \CIV Galactic absorption, requiring pairs to be closer than $3^\circ$. Each of the archival quasars paired to these stars was examined to look for redshifted lines contaminating the Galactic \CIV absorption, and one quasar was rejected for having saturated contamination. From this list we culled out any redundant stars that significantly overlapped, with a bias toward removing fainter stars that would be time-consuming to observe. Finally, we included three sightline pairs from an initial pilot study that made use of preexisting observations of stars \citep{lehner2011} and quasars \citep{zheng2019}. One sightline pair (quasar LBQS-0107-0232 and star J0111-0011) was later excluded from analysis because Lyman-alpha forest prevented accurate measurement of Milky Way absorption features. The final list is composed of 30 BHB star-quasar pairs that provide a roughly even sampling of the high-latitude ($b>30^\circ$) sky (Figure \ref{fig:experimentaldesign}).\\

\subsection{Quasar \& Stellar Spectra}
\label{sec:spectra}

\begin{figure*}[hp!]
    \centering
    \hspace{0.8in}
    \includegraphics{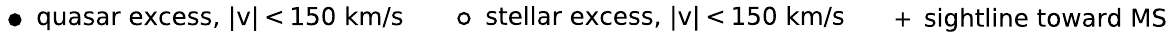}
    \newline
    \vspace{1pt}
    \begin{tabular}{r l}
        \includegraphics[height=7.2in]{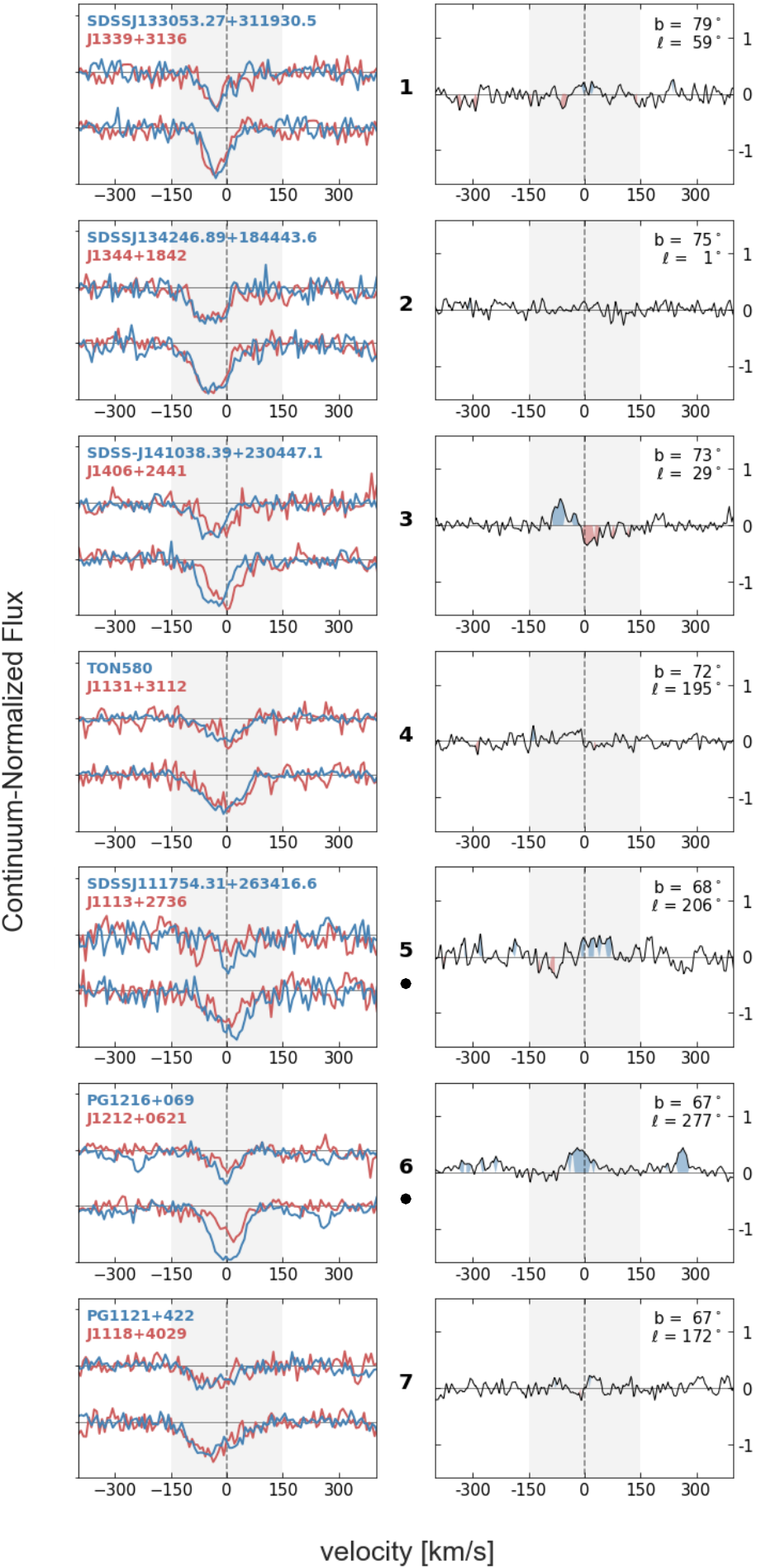}
        \hspace{0.5cm}
        &
        \includegraphics[height=7.2in]{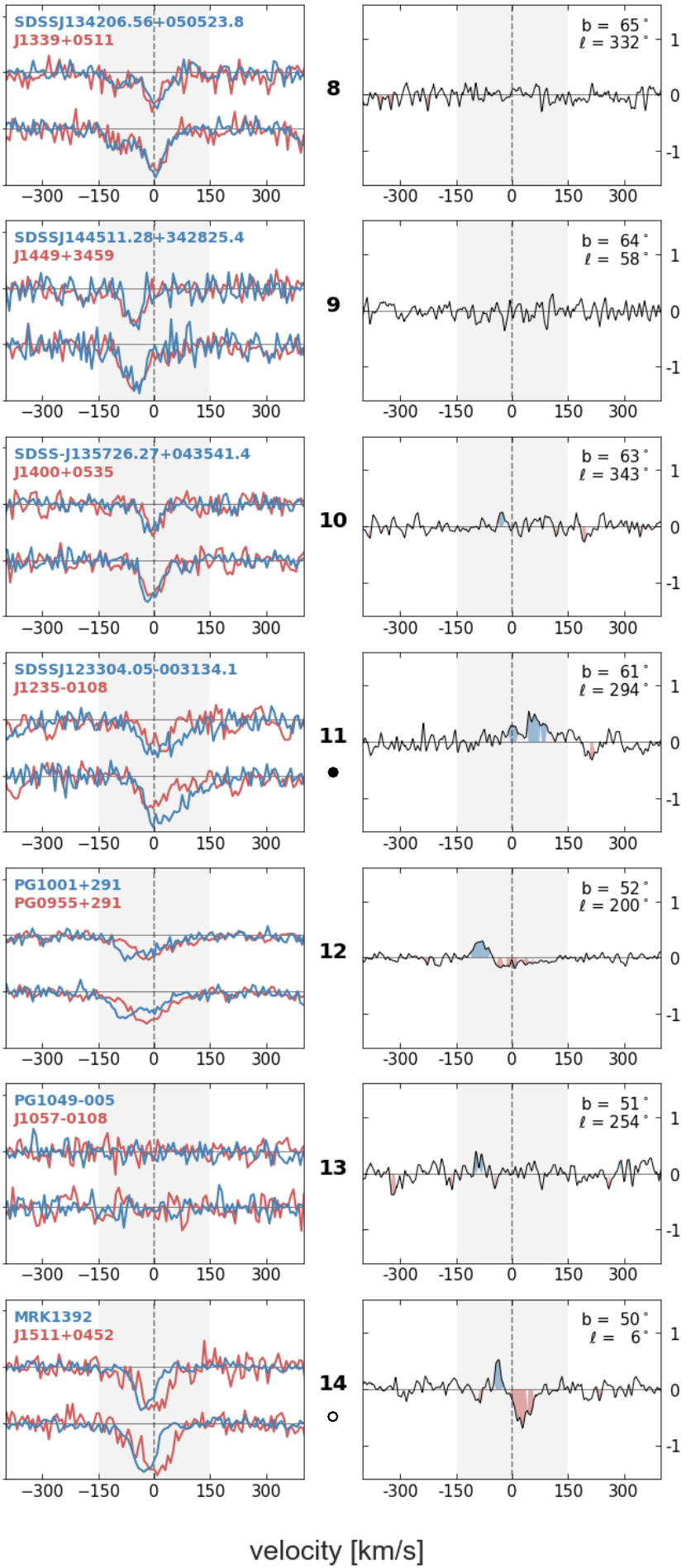}
    \end{tabular}
    \caption{{\sc Difference Spectra.} \textit{Left columns:} \CIV ($\lambda\lambda$  1548, 1550 \AA) absorption doublets in the contamination-corrected, continuum-normalized star-quasar spectrum pairs. Velocities are given in the LSR frame. For each sightline pair, spectra and target names are shown in red for the star and blue for the quasar. The stronger 1548 \AA\, line is shown on bottom with the 1550 \AA\,  line above it. Since the doublet lines are separated by $\sim$500 km/s in velocity space, some features that appear at positive velocities relative to the 1548 \AA\, line are also visible at negative velocities relative to the 1550 \AA\,  line. The sightline pairs are labeled with ID numbers, which are listed in Table \ref{table:properties} and referenced throughout this work. A symbol below the ID number identifies sightline pairs with significant excess quasar absorption~(\QSOexcessdot) or stellar absorption (\BHBexcessdot) at $|v|<150$ km/s. Sightlines in the direction of the MS are labeled with a `+' symbol.
    \textit{Right columns:} Difference spectrum showing the average excess quasar absorption for the doublet lines shown in the left column (quasar absorption - stellar absorption). A value of 0 means that the normalized fluxes of the star and quasar are identical, and a value of $\pm1$ corresponds to an excess equal to the continuum level. In order to highlight features that are inconsistent with noise, regions of the difference spectrum where the excess absorption is greater than the flux error for both lines in the doublet are shaded blue or red for quasar and stellar excess, respectively.
    In both columns, the gray shaded region shows the 150 km/s ``low-velocity'' window within which absorption features were measured.}
    \label{fig:difference-spectra}
\end{figure*}

\setcounter{figure}{1}

\begin{figure*}[hp!b]
    \centering
    \hspace{0.8in}
    \includegraphics{specplots_legend_long.pdf}
    \newline
    \vspace{1pt}
    \begin{tabular}{r l}
        \includegraphics[height=8.2in]{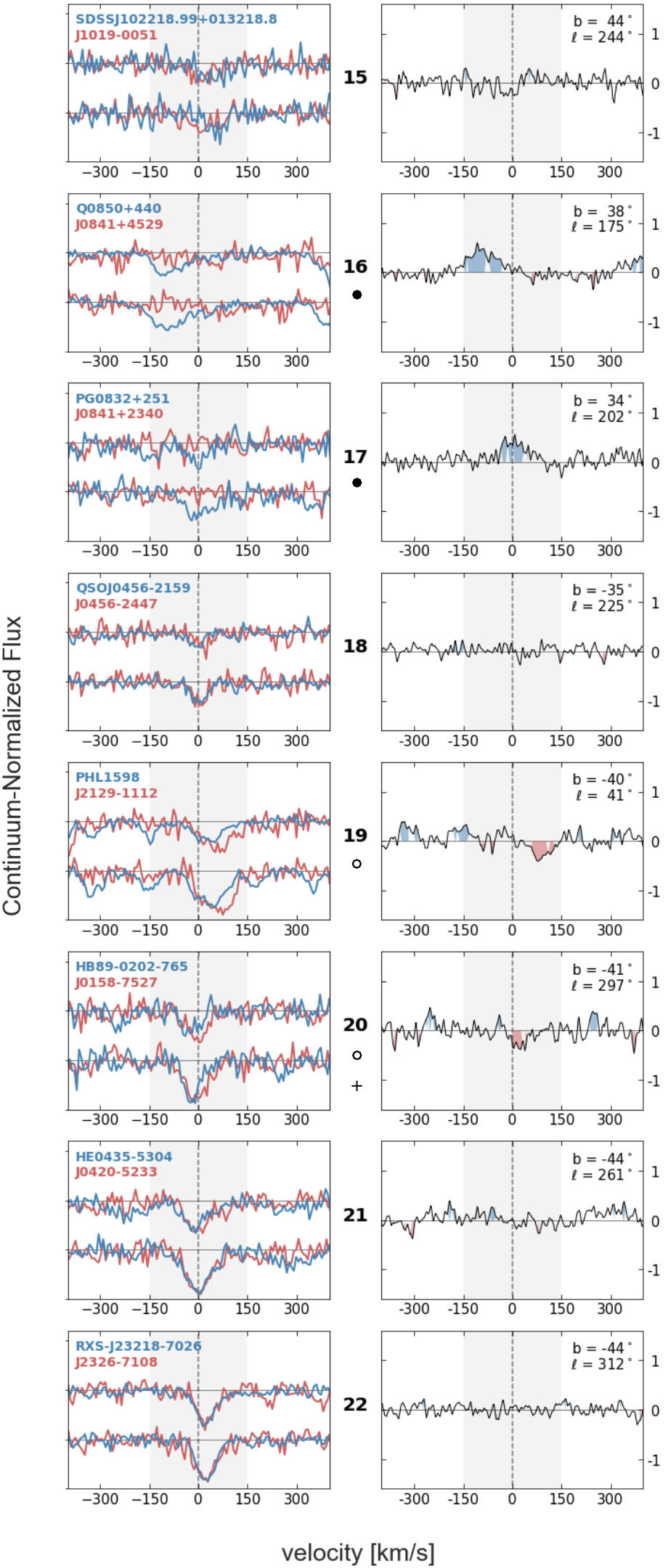}
        \hspace{0.5cm}
        &
        \includegraphics[height=8.2in]{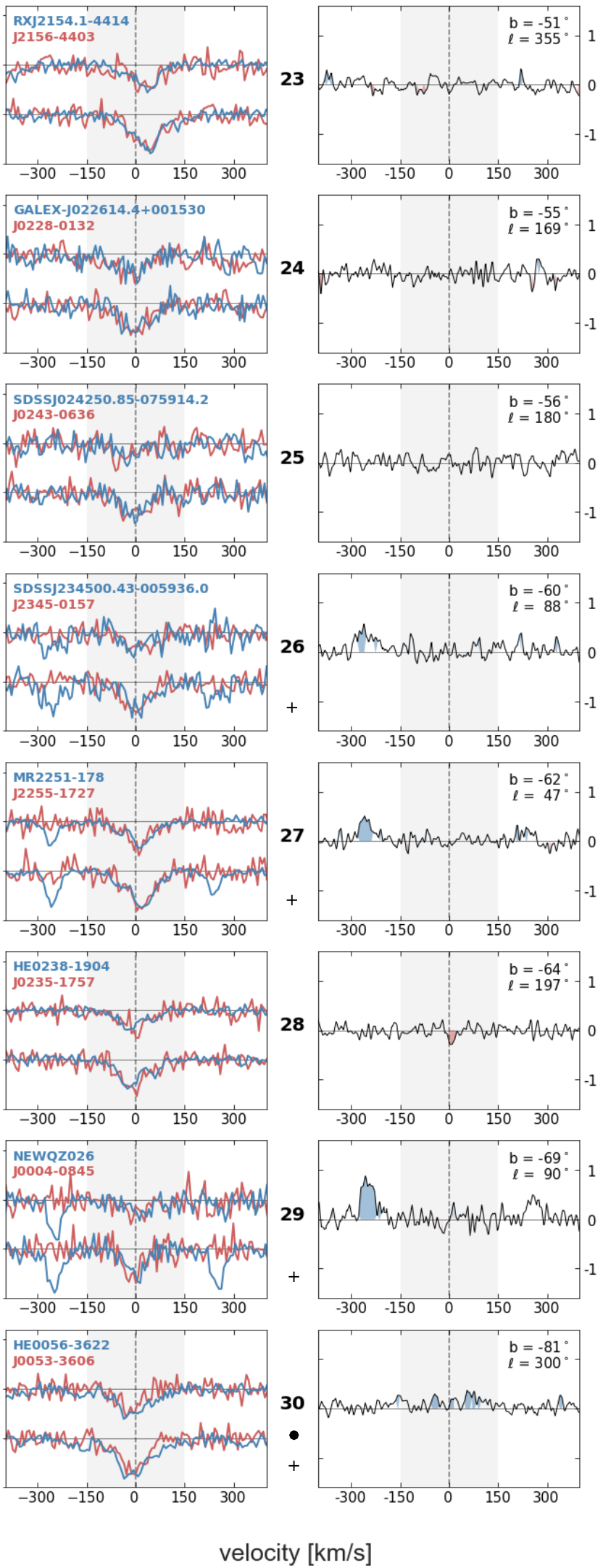}
    \end{tabular}
    \vspace{1cm}
    \caption{(Continued.)}
    \label{fig:difference-spectra-2}
\end{figure*}

\begin{figure*}
    \centering
    \includegraphics[width=\textwidth]{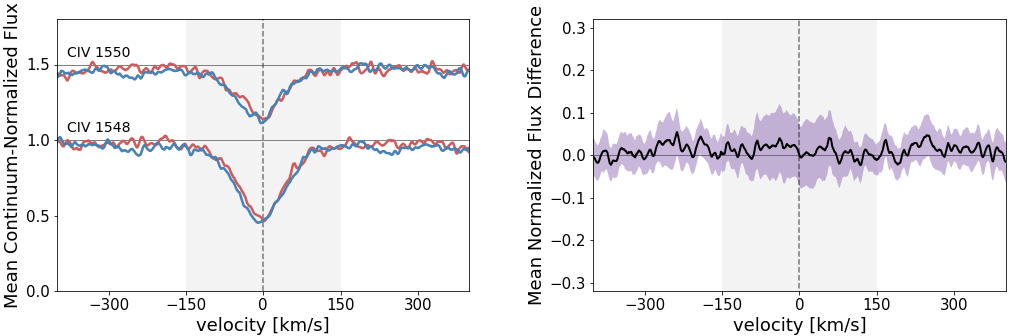}
    \caption{{\sc The bulk of \CIV in the Milky Way sits close to the disk at low velocities.} Mean \CIV absorption profiles (\textit{left}) and difference spectra (\textit{right}) for all 30 individual sightline pairs shown in Figure \ref{fig:difference-spectra}. The low-velocity ($|v|<150$ \kms) measurement window used in this work is shaded gray. The left panel shows the stacked  \CIV ($\lambda\lambda$ 1548, 1550 \AA) doublet profiles for all quasar sightlines in blue and stellar sightlines in red. In the right panel, the corresponding stacked difference spectrum is shown with the standard error on the mean shaded purple. The fluxes are normalized identically in both panels, although the right panel displays a much smaller flux range.\\}
    \label{fig:summed-difference-spectrum}
\end{figure*}

QuaStar includes both stellar and quasar spectra observed with the G160M grating on Cosmic Origins Spectrograph \citep[COS;][]{froning09, green12} on the Hubble Space Telescope. The stellar spectra were obtained as part of a Cycle 26 HST Large Program (PID~\#15656), and the archival quasar spectra were originally obtained as part of a number of different programs with varying science goals.  We detect FUV absorption features from metal ions such \CIV ($\lambda\lambda$ 1548 \AA, 1550 \AA) and \SiIV ($\lambda\lambda$ 1393 \AA, 1402 \AA), which trace warm ionized material. Both our newly observed stellar targets  and their archival quasar pairs in the Hubble Spectral Legacy Archive (HSLA) were observed at a central wavelength of 1577 \AA\,, with a wavelength range of 1383-1754~\AA\, and a velocity resolution of 18 \kms. 

The stellar spectra were co-added with 3-pixel binning using the IDL routine \texttt{COADD\_X1D} (v3.3) developed by \cite{danforth16}. Co-added quasar spectra were obtained directly from the HSLA and rebinned to match the stellar spectra using the SpectRes spectral resampling tool \citep{carnall2017}. 
Coauthor Yong Zheng performed a co-added flux comparison among \texttt{COADD\_X1D} and HSLA data for a set of bright UV stars, finding that despite their slightly different algorithms the spectral co-addition routines from \texttt{COADD\_X1D} and HSLA yield similar line profiles and co-added flux levels (Y. Zheng et al. 2020, private communication). This check assures that the stellar and quasar spectra comparison does not suffer significant systemic errors.
After spectra were binned and co-added, continuum fitting was performed for both stars and quasars using the {\tt linetools} package\footnote{https://github.com/linetools/linetools}, an open-source code for analysis of 1D spectra.

In order to enable accurate measurement of the \CIV column density down to a limit consistent with that of low-redshift CGM surveys (log N$_{\rm CIV}$ $\gtrsim$ 13.5 \percmsq), we obtained COS spectra of stellar sightlines with signal-to-noise ratio (S/N) of $\gtrsim10$ over the full spectrum. The stellar COS spectra have 14.0 $<$ S/N $<$ 43.3 at $\lambda = 1550$ \AA, with a median value of 16.5. The quasar spectra have 5.6 $<$ S/N $<$ 29.5 at $\lambda = 1550$ \AA, with a median value of 14.2. For reference, an S/N of $\sim$15 at $\lambda = 1550$ \AA~ corresponds to a median 2$\sigma$ upper limit (e.g. for non-detections) of log N$_{\rm CIV}$ $\lesssim$ 13.20 \percmsq  over a 50 km/s window. As we will discuss below, our difference spectra require us to measure features over two larger velocity ranges, which in turn impacts our effective sensitivity limit for the column density difference measurement. Specifically, we will consider all material within $\pm$150 km s$^{-1}$ and $\pm$300 km s$^{-1}$ of the local standard of rest (LSR); these windows carry an effective 1$\sigma$ sensitivity limit to $\Delta$ log N$_{\rm CIV}$ of 13.39 and 13.56 cm$^{-2}$, respectively. The sensitivity limits for the difference measurements are explained in more detail in \S\ref{sec:differencespectra}. \\

\section{Analysis}
\label{sec:analysis}

\subsection{Contamination Correction}
\label{sec:contaminationcorrection}

Once data reduction was complete, we identified and corrected for contaminating absorption features, which were present in 16 quasar spectra and 17 stellar spectra. In quasar spectra, where absorption-line systems at higher redshifts can overlap with rest-wavelength features of interest, there were a greater number of contamination features per sightline and those features had larger column densities on average. Contaminating absorption lines intrinsic to the BHB stellar atmospheres were less numerous, and when present, the contaminating absorption was weaker. 

In order to distinguish contaminants from Galactic absorption, we used the {\tt PyIGM IGMGUESSES} GUI\footnote{https://github.com/pyigm/pyigm} to identify all absorption features within 500 km/s of the \CIV ($\lambda\lambda$  1548, 1550 \AA) and \SiIV ($\lambda\lambda$ 1393, 1402 \AA) doublets and obtained preliminary line profile fits (with velocity $v$, column density $N$, and Doppler b parameter $b_{\rm D}$). 
We found that absorption features in the \CIV doublet consistently matched one another but \SiIV features were frequently saturated, resulting in unreliable column density measurements. For that reason, we focus only on \CIV measurements in the body of this paper.

To remove contaminating absorption, the velocity window used to fit each absorption component was determined by eye, and both lines in each absorption doublet were then fit simultaneously. This individual inspection allowed us to assign narrow velocity priors for fitting in cases when multiple components were blended. Any remaining absorption features that were not present in both lines of the \CIV doublet were considered contaminants. If we found any blended contaminants in quasar spectra, we then identified and fit profiles for all absorption-line systems at higher redshifts to identify the contaminating line. Once all absorption components were identified and fit, we adjusted the normalized fluxes of each spectrum using the best-fit line profiles of the contaminants. All subsequent analysis was performed on the resulting contamination-corrected spectra.\\

\subsection{Column Densities}
\label{sec:columndensities}

We measured column densities using the apparent optical depth method (AODM) with the same {\tt linetools XSpectrum1D} package\footnote{https://github.com/linetools/linetools} used for data reduction (see \S\ref{sec:spectra}). We measured the \CIV ($\lambda\lambda$  1548, 1550 \AA) and \SiIV ($\lambda\lambda$ 1393, 1402 \AA) doublets, which are both strong and well-defined features in UV COS spectra. However, saturated \SiIV features often prevented precise or reliable measurement, and we therefore present only the \CIV measurements in this work.

We report column densities measured for two velocity windows, which include absorption within 150 and 300 km/s of the Galactic LSR for each sightline. 
In our primary analysis we focus on the $|v|<150$ km/s window. We choose this velocity range in order to fully capture the low-velocity absorption features we detect while minimizing the potential for including additional noise or contaminating features. Going forward we use the term ``low-velocity"  to refer to this $|v|<150$ km/s window unless otherwise specified.
In addition, we make the same measurement within a wider $|v|<300$ km/s window, which includes absorption from the MS along some quasar sightlines. If structures similar to the MS exist in external galaxies, then they could potentially contribute to CGM column densities measured at low velocity relative to the galaxy, depending on the sightline orientation and corresponding observed radial velocity. We therefore find this wider velocity window useful for comparing our results to column densities of external galaxies in the literature (see \S\ref{sec:litcomparison} for such comparisons).\\

\subsection{Difference Spectra}
\label{sec:differencespectra}

In order to isolate and measure absorption signatures of CGM gas beyond the disk-halo interface, which is always blended with absorption from foreground gas along the line of sight, we leverage the unique experimental design of QuaStar to correct for the blended foreground absorption. Data included in the archival COS-GAL sample indicate that \CIV absorption exhibits no detectable variation between $\sim80\%$ of sightlines separated by $<2^\circ$ \citep{zheng2019}; we can reasonably assume that the \CIV absorption originating from foreground gas will be consistent between the stellar and quasar spectra in the same fraction of our close sightline pairs (but see \OVI analysis of SMC/LMC sightlines of Howk et al. 2002\nocite{Howk2002}). Therefore, we can measure the absorption from foreground gas along stellar sightlines that extend to the disk-halo interface and use those measurements to subtract foreground gas absorption along the closely paired quasar sightlines. We expect that $\sim$20\% of paired sightlines will have excess stellar or quasar absorption due to clumpy foreground gas. In theory, the resulting difference spectrum will reveal any remaining \CIV absorption originating from Milky Way CGM gas behind the star. We discuss some caveats to these assumptions in \S\ref{sec:litcomparison}.

Figure \ref{fig:difference-spectra} displays line profiles of the \CIV ($\lambda\lambda$  1548, 1550 \AA) absorption doublet for each paired star (red) and quasar (blue), along with the corresponding difference spectrum for each sightline pair. Each pair is labeled with an ID number that can be referenced in Table \ref{table:properties}, and a symbol appears below the ID number in cases where the sightline pair has significant excess quasar absorption~(\QSOexcessdot) or stellar absorption~(\BHBexcessdot) at $|v|<150$ km/s. Sightlines in the direction of the Magellanic System are labeled with a `+' symbol, regardless of whether excess absorption is detected. The spectra (left columns) are continuum normalized and corrected for contamination, and the corresponding difference spectra (right columns) are determined simply by subtracting the stellar flux from the quasar flux. Thus, the difference spectra are positive where absorption along the quasar sightline exceeds absorption along the stellar sightline, and vice versa. The difference spectra in Figure \ref{fig:difference-spectra} are shaded where this excess is greater than flux measurement errors, with red shading used for excess stellar absorption and blue for excess quasar absorption.

The mean stacked absorption profile and difference spectrum of the individual sightline pairs in Figure \ref{fig:difference-spectra} are presented in Figure \ref{fig:summed-difference-spectrum}. As with the individual sightline pairs, the left panel shows the \CIV absorption doublet, with the mean normalized flux for quasars in blue and for stars in red. The right panel shows the mean stacked difference spectrum for the 30 sightline pairs. The standard error on the mean, shaded purple, is calculated using the jackknife resampling method, which does not assume a normal distribution or large sample size. The absorption profile in the left panel illustrates that the bulk of the \CIV column density we detect in the Milky Way is moving at velocities within 100 \kms of the LSR. However, the difference spectrum in the right panel is relatively flat and consistent with zero, which suggests that little of the low-velocity \CIV lies beyond $d\approx14$ kpc in the Milky Way's halo. While stacked spectra are useful for getting a sense of global behavior over the whole sky, we caution that they may be misleading. Features of interest that are weak and present in only one direction on the sky may be smoothed out. In a forthcoming paper we will further examine such features and the kinematic information they provide about Galactic features such as rotational lag and gas flows.

If foreground absorption were perfectly homogeneous and identical along both quasar and stellar sightlines in every case, we would not expect to see any stellar excess in the difference spectra. However, as discussed in \S\ref{sec:observations}, we expect that variation in foreground gas density due to clumpy ISM substructure will result in a significant detection of excess absorption for 20\% of our paired star and quasar sightlines. In our sample of 30 sightline pairs, this translates to 3$\pm$1 pairs with detectable stellar excess and 3$\pm$1 pairs with detectable quasar excess, assuming Poisson noise. We will report both an empirical covering fraction, uncorrected for this clumpiness factor, and a corrected covering fraction, which excludes three detections assumed to be contamination. 

To determine our sensitivity to weak CIV features in the difference spectra, we estimate upper limits on N$_{\rm CIV}$ for each spectrum using the 1$\sigma$ error on column density measured within a featureless region of the continuum near the \CIV ($\lambda\lambda$  1548, 1550 \AA) doublet. The velocity width of the featureless region matches the width of the window used for column density measurements, but the exact location of the featureless region in velocity space was selected by eye to ensure that contaminating lines were excluded from the measurement. These 1$\sigma$ detection limits are conservative upper limits given the large velocity ranges over which they are measured. The upper limit we report on the difference measurement $\Delta$logN$_{\rm CIV}$ in the case of a non-detection, $N_{det}$, is the two one-sigma upper limits for stellar and quasar sightlines added in quadrature.\\

\section{Results}
\label{sec:results}

\begin{figure*}
    \includegraphics[width=1.\textwidth]{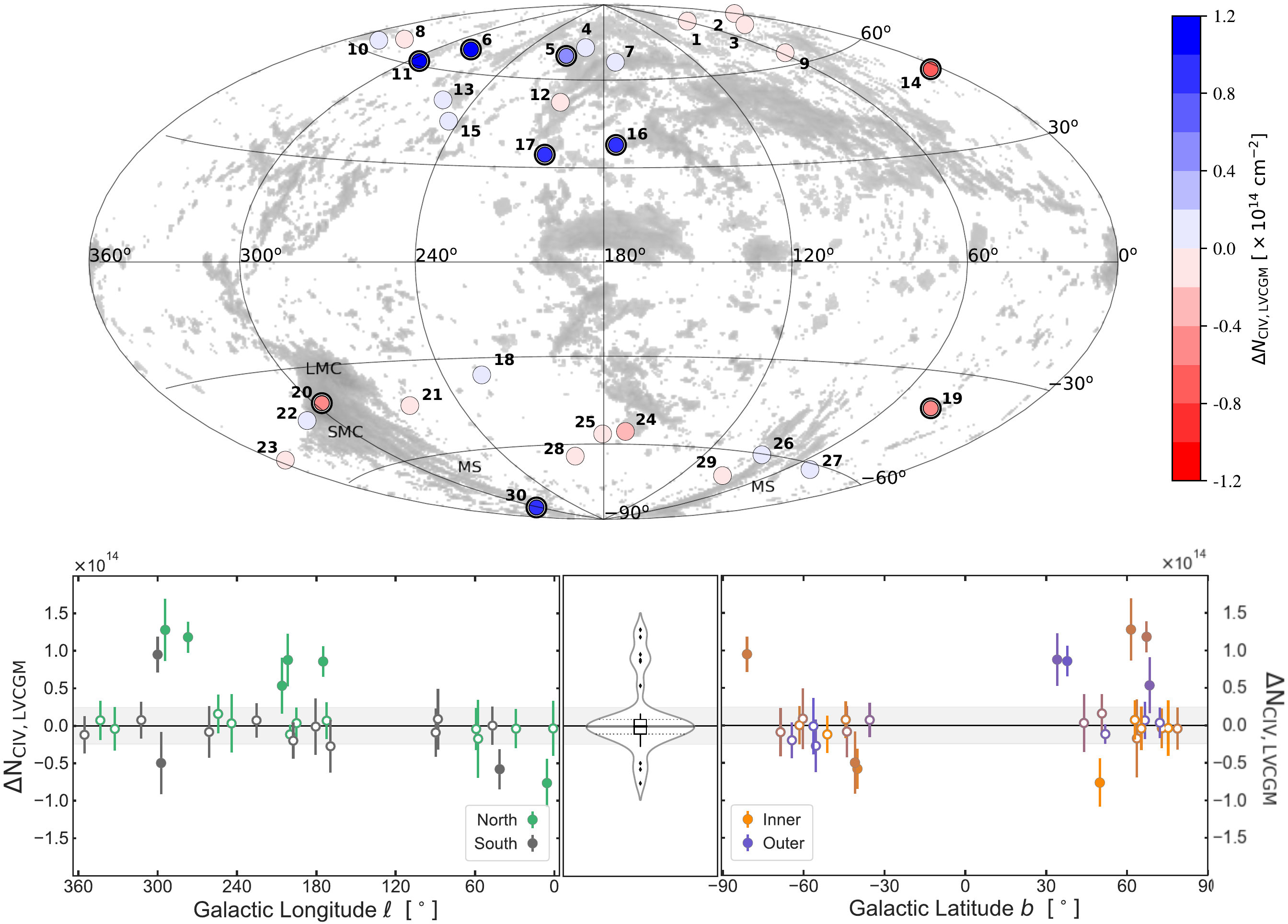}
    \caption{{\sc Low-velocity $\CIV$ in the Milky Way's CGM.} \textit{Top:} Full-sky map of Galactic high-velocity clouds from the HI4PI survey in gray \citep{bekhti2016,westmeier2018} overlaid with QuaStar column densities for low-velocity halo gas. Specifically, \diffLVCGM is the difference in low-velocity ($|v| < 150$ km/s) \CIV gas column density between the star and quasar in each sightline pair. Each marker represents one star-quasar sightline pair. Blue markers/positive values indicate an absorption excess along the quasar sightline, which likely originates from the Milky Way's extended CGM. Red markers/negative values indicate absorption excess along the stellar sightline, which is likely due to small-scale foreground fluctuations in \CIV column density. Thick marker edges highlight sightline pairs with significant absorption excess. The Magellanic Clouds and Magellanic Stream are labeled for reference. \textit{Bottom:} Column density difference of sightline pairs vs. Galactic longitude and latitude. Here, color represents the galactic coordinates of the sightline pair. Open markers represent measurements below the sensitivity limit. In the left panel, markers are green if a sightline is in the northern Galactic hemisphere and gray if it is in the southern Galactic hemisphere; in the right panel, markers are orange if the sightline is toward Galactic center, and purple if it is away from Galactic center. The violin plot in the middle panel shows the distribution, median value, and interquartile range of the \CIV column density difference measurement \diffLVCGM.  The range of column density measurements that fall below the median sensitivity limit of the sample (\diffLVCGM$<13.39$ cm$^{-2}$) is shaded gray. Notably, the column density difference \diffLVCGM for most sightline pairs falls below the sensitivity limit, suggesting that there is little low-velocity gas in the Milky Way's extended CGM.\\}
    \label{fig:diffplots-combined}
\end{figure*}

We measure \CIV column densities for each individual sightline, and the difference between those column densities for each star-quasar sightline pair. For convenience we define our column density difference measurement of low-velocity \CIV in the Milky Way's CGM to be
\begin{align*}
    \Delta logN_{LVCGM}& 
    \begin{cases}
        = log(N_{q}-N_{s}) \hfill & \\
        \qquad\qquad\qquad if\ \hspace{2pt}\QSOexcessdot[10]:\ \,\hspace{6pt}  |N_{q}-N_{s}|&\hspace{-7pt}>\hspace{-2pt} N_{det} \\
        \vspace{7pt}
        \hfill N_{q} &\hspace{-7pt}>\hspace{-2pt}N_{s}\\
        \vspace{2pt}
        = -log(N_{s}-N_{q}) \hfill & \\
        \qquad\qquad\qquad if\ \hspace{-0.5pt}\hspace{2pt}\BHBexcessdot[16]:\ \,\hspace{6pt}  |N_{q}-N_{s}| &\hspace{-7pt}>\hspace{-2pt} N_{det} \\
        \vspace{7pt}
        \hfill N_{q} &\hspace{-7pt}<\hspace{-2pt}N_{s}\\
        \vspace{2pt}
        < N_{det} \hfill & \\
        \qquad\qquad\qquad if\ \hspace{2pt}\varnothing:\ \,\hspace{6pt}  |N_{q}-N_{s}| &\hspace{-7pt}<\hspace{-2pt} N_{det}
    \end{cases}
\end{align*} 
where $N_q$ is the quasar sightline column density, $N_s$ is the stellar sightline column density, and the difference measurement sensitivity limit $N_{det}$ is the sensitivity limit of the stellar and quasar spectra added in quadrature, as discussed in \S\ref{sec:differencespectra}. Thus, \diffLVCGM is the log of the difference in column density for a star-quasar pair, where excess absorption along the quasar sightline gives a positive value and excess absorption along the stellar sightline gives a negative value. If the column density difference is smaller than the sensitivity limit, then $N_{det}$ places an upper limit on the value of \diffLVCGM.   
The difference measurement \diffLVCGM  will be negative in cases where there is a small clump or overdensity present along the line of sight to the star that is not in the foreground gas along the quasar sightline. We expect roughly 10\% of sightline pairs to have column density excess along the stellar sightline due to differences in foreground gas absorption (see \S\ref{sec:differencespectra}).

We present the results for two velocity windows: one measuring $|v|<150$ km/s gas (hereafter referred to as ``low-velocity"), and a wider $|v|<300$ km s$^{-1}$ window that is useful for comparing to extragalactic CGM measurements (see \S\ref{sec:columndensities} for an explanation of this choice of velocity windows). The low-velocity ($|v|<150$ km/s) \CIV column densities and difference measurements for each star-quasar sightline pair are listed in Table \ref{table:properties}. Low-velocity \CIV column densities for stellar sightlines have a range of logN$_{\rm CIV, star}$~$=$~$[13.12,14.41]$ and the log of the median column density is 14.11;  quasar sightlines have a range of logN$_{\rm CIV, quasar}$~$=$~$[13.47,14.40]$ and the log of the median column density is 14.16. The column density difference measurements for sightline pairs have a range of \diffLVCGM~$=$~$[11.41,14.11]$ for pairs with excess quasar absorption and \diffLVCGM~$=$~$-[12.03,13.88]$ for pairs with excess stellar absorption. Over all \diffLVCGM measurements, the log of the median column density difference is $-11.60$ and the median sensitivity limit is $logN_{det}=13.39$. The distribution of these values is shown by the violin plot at the center of the bottom panel of Figure \ref{fig:diffplots-combined}. Within the low-velocity window we detect excess stellar absorption above the sensitivity limit for three sightline pairs (IDs 14, 19, \& 20) and excess quasar absorption for six sightline pairs (IDs 5, 6, 11, 16, 17, \& 30). We find a total low-velocity \CIV covering fraction of $f_{{\rm CIV,} star}$(log N$>$13.65)=97\% [29/30] and $f_{{\rm CIV,} quasar}$(log N$>$13.65) = 100\% [30/30]; after subtracting foreground gas absorption with the spectral differencing technique, the \CIV covering fraction of the isolated CGM component is $f_{\rm CIV, LVCGM}$(log N$>$13.65)=20\% [6/30]. If we assume that three of these sightlines are false positives caused by clumpy foreground gas (as discussed in \S\ref{sec:differencespectra}), then the corrected covering fraction is $f^*_{\rm CIV, LVCGM}$(log N$>$13.65)=10\% [3/30]. To test our assumption that there is negligible variation of \CIV~ column density on scales of $<$2$^{\circ}$, we confirmed that $|$\diffLVCGM$|$ is not correlated with sightline separation of the star-quasar pair (p=0.67). We also confirmed that \diffLVCGM is not correlated with Galactic latitude (p=0.58), Galactic longitude (p=0.37), or distance to the foreground star (p=0.71).

We repeat these measurements within the wider $|v|<300$ km/s velocity window and obtain difference measurements with a range of \diffLVCGM~$=$~$[12.04,14.28]$ for pairs with excess quasar absorption and \diffLVCGM~$=$~$-[12.42.13.92]$ for pairs with excess stellar absorption, with a log of the overall median column density difference of 12.65 and an overall median sensitivity limit of $logN_{det}=13.56$. Within this wider velocity window we detect excess stellar absorption above the sensitivity limit for 2 sightline pairs (IDs 14 \& 19), and excess quasar absorption for 10 sightline pairs (IDs 5, 6, 11, 16, 17, 21, 26, 27, 29, \& 30). The covering fraction of CGM gas within the wider velocity window is $f_{\rm CIV, CGM}$(log N$>$13.65)=38\% [9/24], and the corrected covering fraction is $f_{\rm CIV, CGM}$(log N$>$13.65)=25\% [6/24]. Note that this covering fraction excludes any sightline pairs with a sensitivity limit greater than the given limiting column density. The absorption features we detect at $150<|v|<300$ km/s are typically distinct from any excess absorption at $|v|<150$ km/s and do not suffer from blending. We identify the MS as the absorber in most of these sightlines and discuss possible origins of other excess absorption in \S\ref{sec:MSandLVHCs}.

The top panel of Figure \ref{fig:diffplots-combined} shows the \diffLVCGM difference measurements within a $|v|<150$ km/s velocity window, overlaid on a map of Galactic HVCs ($|v| > 70$ \kms) from the HI4PI survey in gray \citep{bekhti2016,westmeier2018}. Each marker represents one star-quasar sightline pair and is labeled with an ID that can be referenced in Table \ref{table:properties} and Figure \ref{fig:difference-spectra}. Markers are blue if there is more \CIV absorption along the quasar sightline, and red if there is more along the stellar sightline. The bottom panel plots the same difference measurements against Galactic latitude and longitude separately, with their distribution in the middle. Each marker again represents one sightline pair, but here the color of the markers corresponds to their Galactic coordinates. The median sensitivity limit for the sample is shaded gray for reference, but each sightline pair may have a sensitivity limit above or below this value.

The fraction of sightline pairs with significant excess stellar absorption (3/30) is consistent with our expectation that 10\% of sightline pairs will have stellar excess due to clumpiness in foreground gas (see \S\ref{sec:differencespectra}). The quasar excess detection rate (6/30) suggests that the column density of extended Milky Way CGM gas is below our sensitivity limit of logN$_{\CIV}<13.39$ \percmsq across most of the sky. All but one of the significant detections are in the northern Galactic hemisphere at latitudes $60^{\circ}<b<70^{\circ}$. The two sightline pairs with the greatest excess quasar absorption (IDs 6 \& 11) are close together at Galactic longitudes $270^{\circ}<\ell<300^{\circ}$.\\

\section{Discussion}
\label{sec:discussion}

In this section we comment on the significance of specific difference measurement detections and identify the MS, a low-velocity cloud, and HI Complex A as possible origins of the detected absorption. We then compare our observations of the Milky Way's CGM to similar measurements of external galaxies in order to place our results in larger context. Finally, we discuss apparent inconsistencies between the Milky Way and other galaxies, and explore some explanations which might reconcile them.\\

\subsection{The Magellanic System and LVHCs}
\label{sec:MSandLVHCs}

One of the most prominent features visible in the difference spectra is the excess quasar sightline absorption at $v\sim-250$~km/s for pair IDs 26, 27, and 29 (see Figure \ref{fig:difference-spectra}). The velocities and Galactic coordinates of these features are consistent with absorption from the Magellanic Stream \citep{nidever2008,fox2014} and they have \CIV column densities logN$_{\rm CIV} =$ 13.81, 13.78, and 14.16, respectively. Those three prominent absorbers account for 96\% of the excess absorption added by the wider velocity window (that is, at $150 < |v| < 300$ km/s), which increases the median difference measurement of the overall sample by $\Delta$logN=12.46  from the narrower low-velocity window. Two other sightlines also have absorption identified as MS because they are in the direction of MS-associated \HI gas: sightline 30 intersects a section of the Magellanic Stream with low radial velocities and shows some detectable absorption within the low-velocity window, while sightline 20 lies in the direction of the Magellanic Clouds and has a stellar excess within the low-velocity window.

The considerable contribution of high-velocity gas to the total column density detected in the Milky Way's CGM is an apparent contradiction to extragalactic CGM studies that detect $>90\%$ of both low- and high-ionization CGM gas at velocities within 200 km/s of the galaxy systemic velocity \citep[e.g.,][]{tumlinson2013,werk2016}. But the MS is a dominant feature peculiar to the Milky Way and therefore presents a complication when interpreting our results in the context of other Milky Way-like galaxies.

Our primary analysis of low-velocity gas, however, focuses on a $|v| < 150$ km/s window that excludes most MS absorption. Within that narrower window we detect low-velocity CGM gas for six sightline pairs. Three of these detections are likely due to expected fluctuations of foreground gas density (see \S\ref{sec:analysis}). Furthermore, the quasar excess for pair \#16 may be capturing the ionized edge of Complex A, which has a distance of 8 kpc $\lesssim d\lesssim10$ kpc \citep{wakkervanwoerden1997,wakker2001,sembach2003}. 
The two remaining detections have a clear correlation with known structures. Sightline pairs 6 and 11 have the largest \diffLVCGM measurements, and both lie in the northern Galactic hemisphere in the direction of a HI low-velocity halo cloud (LVHC) previously identified by \cite{Peek2009a}. 

Low-velocity clouds represent a distinct population that has not been well studied, but they are associated with detections of both HI and metals that trace cooler gas, typically in the lower halo or the Milky Way disk \citep{bekhti2009}. The low-velocity cloud that corresponds to our detections is a continuation of a large HVC complex in the region \citep{Peek2009a, saul2012}; aside from the MS, all known HVC complexes have distances that place them in the inner halo ($d\lesssim15$ kpc). It is therefore unlikely that the low-velocity gas we detect in sightline pairs 6 and 11 originates in the extended CGM. 

Finally, we note that five of the six quasar excess detections are found in the northern Galactic hemisphere. Although we do not know the distance to any of these absorbers with certainty, their distribution suggests that low-velocity CGM gas in the Milky Way is not spatially homogeneous.\\

\begin{figure*}
    \centering
    \includegraphics[width=0.95\textwidth]{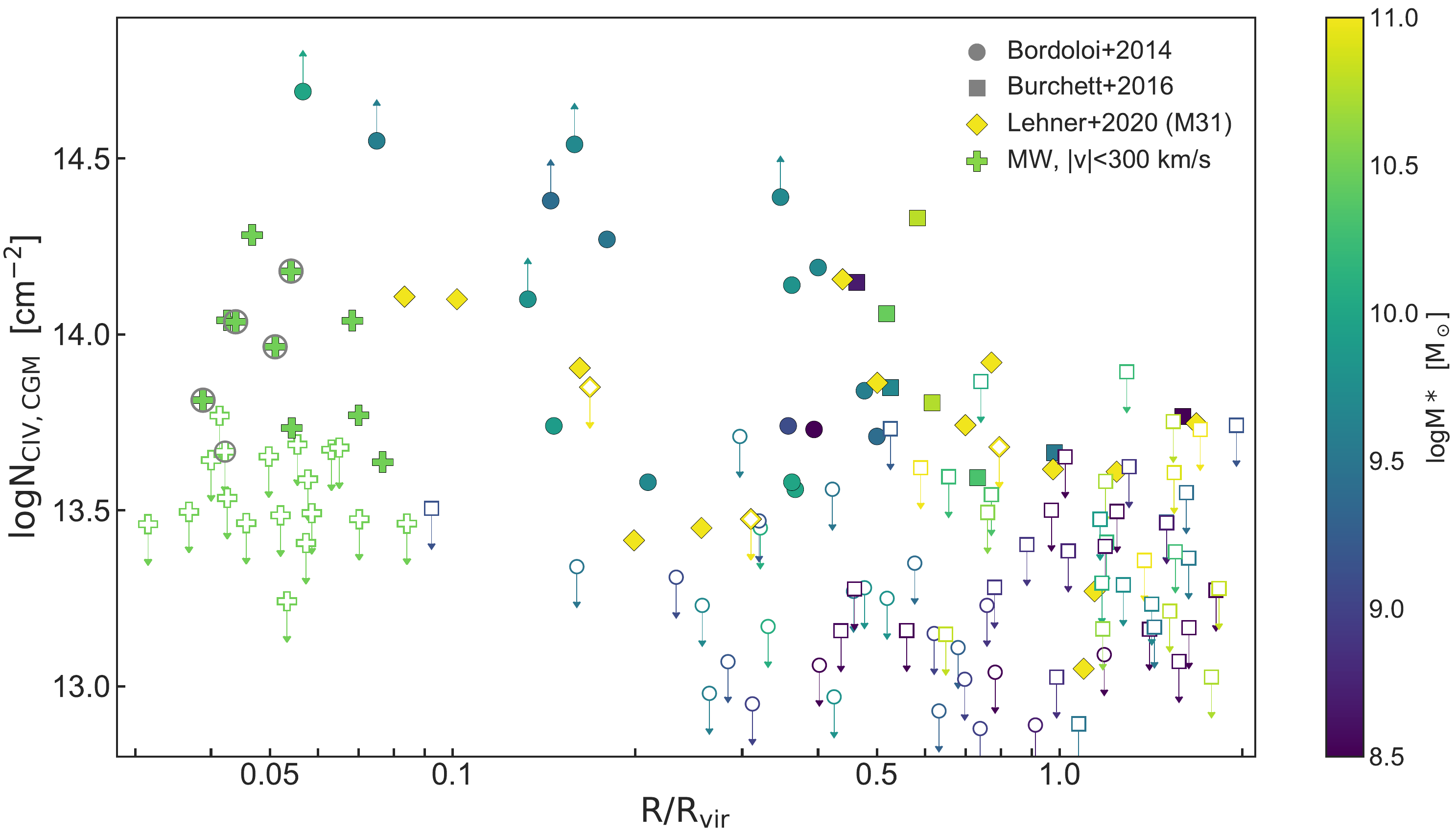}
    \caption{{\sc CGM column density in Milky Way vs. other galaxies.} Column densities of \CIV in the CGM of the Milky Way and other low-redshift galaxies \citep{Bordoloi2014, Burchett2016, lehner2020} as a function of distance from the center of the host galaxy normalized by virial radius. In the case of external galaxies, $R/R_{vir}$ is the impact parameter $\rho$ of the sightline normalized by the galaxy's virial radius. For QuaStar sightlines, which have a different viewing geometry than extragalactic sightlines, $R/R_{vir}$ is the paired star's Galactocentric distance normalized by the Milky Way's virial radius. The markers are filled if the measurement is a detection; otherwise, the upper limit is shown with an open marker. Marker colors represent the stellar mass of the galaxy being probed by the sightline. The Milky Way column density difference measurements in QuaStar, represented by `+' symbols, are measured within a $|v|<300$ km/s window, which includes the MS. QuaStar sightlines in the direction of the MS are circled.\\}
    \label{fig:litcomparison}
\end{figure*}

\subsection{Comparison with Low-Redshift CGM Surveys}
\label{sec:litcomparison}

In \S\ref{sec:results} we reported excess quasar absorption in 6/30 sightline pairs, and we expect that $3\pm1$ of those detections are due to fluctuations of foreground gas (see \S\ref{sec:differencespectra}). Two of the remaining three detections originate in a low-velocity HI halo cloud that is likely to be located in the inner Galactic halo. Within a wider velocity window, almost all detected absorption originates from the MS. In sum, we detect very little low-velocity gas in the Milky Way's extended CGM. Are these results expected based on what we know about low-velocity halo gas in external Milky Way-like galaxies? In order to understand our findings, we must examine them in the context of existing extragalactic observations.

In this section we compare QuaStar column density measurements and covering fractions of \CIV with similar measurements in external low-redshift galaxies. In addition to the low-velocity ($|v| < 150$ km/s) window used throughout our main analysis and discussion of this work (Figure \ref{fig:diffplots-combined} and Table \ref{table:properties}), we use a wider velocity window ($|v| < 300$ km/s) to measure column density difference. This expanded velocity window is useful for comparison with external galaxies, which often include a velocity search window of $\pm$300 km s$^{-1}$ from the galaxy redshift. However, complication arises with comparisons using the expanded velocity window because it includes MS absorption in the Milky Way, whereas there is scant evidence for LMC-like companions in other galaxies \citep{tollerud2011, liu2011, jamesivory2011}. For that reason we consider measurements within the wider velocity window to be more conservative in the context of investigating whether the Milky Way has anomalously sparse low-velocity CGM compared to other galaxies. 

In Figure \ref{fig:litcomparison}, we plot column densities of \CIV as a function of $R/R_{vir}$, the distance from the center of the host galaxy normalized by its virial radius. In the case of external galaxies, $R/R_{vir}$ is the impact parameter $\rho$ of the sightline normalized by the galaxy's virial radius. For QuaStar sightlines, which have a different viewing geometry than extragalactic sightlines, $R/R_{vir}$ is the paired star's Galactocentric distance normalized by the Milky Way's virial radius. Column densities are represented by markers colored according to the stellar mass of the host galaxy; the markers are filled if the measurement is a detection; otherwise, the upper limit is shown with an open marker. The Milky Way column density difference measurements in QuaStar, represented by `+' symbols, are measured within a $|v|<300$ km/s window that includes the MS. QuaStar sightlines in the direction of the MS are circled. We assume a virial radius of 230 kpc for the Milky Way, determined by adopting a mass of $M_{vir} = 1.3\times 10^{12}$ \citep{postihelmi2019}, $\rho_{crit} = 2.78\times10^{11} h^2 M_\odot$ Mpc$^{-3}$ \citep{planck2013}, and the $R_{200}-M_{halo}$ relation, $R^3_{200}=3M_{halo}/4\pi\Delta_{vir}\rho_{matter}$. 

We compare our results with \CIV column densities of low-velocity gas in the halos of M31 \citep{lehner2020}, a sample of nearby galaxies \citep[$z<0.015$;][]{Burchett2016}, and the COS-Dwarfs survey \citep[$z<0.1$;][]{Bordoloi2014} in Figure \ref{fig:litcomparison}. It should be noted that many of these galaxies have stellar masses significantly lower than the Milky Way and do not necessarily reflect the distribution of nearby galaxies. The CGM of dwarf galaxies may have a different density profile than \lstar{} galaxies, and they are thought to experience halo-scale flows tied to star-formation activity that may drive gas loss \citep{oppenheimer2018b, johnson2017, Li2021}. Ideally we would only compare our results to measurements of galaxies with masses and star-formation rates similar to the Milky Way, but the data for \CIV in galaxy halos are sparse. At this time, the column densities we show in Figure \ref{fig:litcomparison} represent the only substantial datasets available with \CIV column densities around low-redshift galaxies.

The available measurements of \CIV around external galaxies exhibit a trend of increasing column density at smaller impact parameters, and at $R/R_{vir} < 0.2$ most log$\rm N_{CIV}$ measurements fall well above the sensitivity limit of QuaStar (the majority are lower limits as a result of saturation). The few Milky Way CGM detections are slightly lower than expected based on this trend, although they appear roughly consistent with the \CIV in M31 as measured by the AMIGA survey \citep{lehner2020}. However, most QuaStar sightlines do not show detectable \CIV in the Milky Way's CGM; this is inconsistent with the external sightline measurements, particularly at low impact parameters.

\begin{figure}
    \centering
    \includegraphics[width=0.49\textwidth]{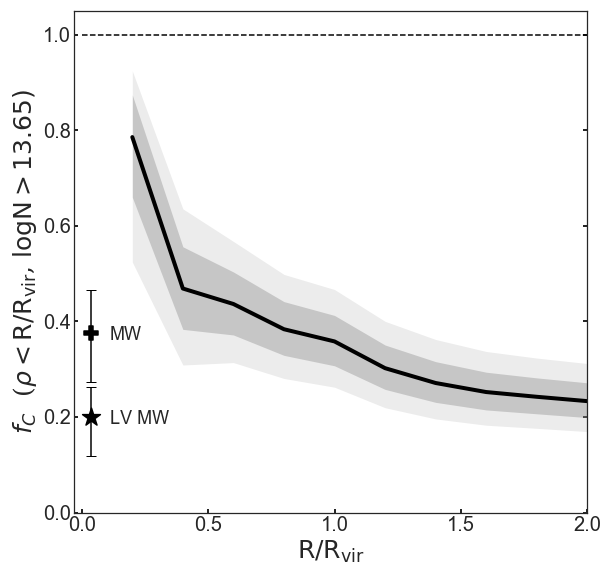}
    \caption{{\sc CGM covering fraction in Milky Way vs. other galaxies.} Cumulative covering fraction $f_C$ of low-velocity $\CIV$ CGM gas for impact parameters within $R/R_{vir}$ of the host galaxy and a limiting column density of 13.65. The black line represents the covering fraction of low-velocity CGM gas for the combined samples of external galaxies presented in Figure \ref{fig:litcomparison} \citep{Burchett2016, Bordoloi2014, lehner2020}. The 1$\sigma$ and 2$\sigma$ Wilson binomial confidence intervals are shaded light and dark gray, respectively. The star symbol represents the covering fraction and 1$\sigma$ confidence interval for low-velocity CGM detections in QuaStar (``LV MW'', $|v|<150$ km/s), and the plus symbol represents the same measurement within a wider velocity window that includes the MS (``MW'', $|v|<300$ km/s). The cumulative covering fraction for external galaxies grows larger at smaller impact parameters, approaching unity at $R/R_{vir}<0.1$. The CGM covering fraction in the Milky Way, as observed from within the Galaxy at $R\sim$ 0.03 $R_{vir}$, is inconsistent with the trend for the CGM covering fractions in external galaxies.}
    \label{fig:coveringfractions}
\end{figure}

The striking discrepancy between the Milky Way and other low-redshift galaxies is illustrated most clearly in Figure \ref{fig:coveringfractions}. The black line represents the empirical cumulative covering fraction for all sightlines within some radial distance of a galaxy, calculated for the three data sets in Figure \ref{fig:litcomparison} combined \citep{Bordoloi2014, Burchett2016, lehner2020}. The 1$\sigma$ and 2$\sigma$ Wilson binomial confidence intervals are shaded light and dark gray, respectively. The symbols with error bars represent the covering fraction and 1$\sigma$ confidence interval for Milky Way CGM detections in QuaStar within the low-velocity window (``LV MW'', $|v|<150$ km/s) and broader window that includes MS absorption (``MW'', $|v|<300$ km/s). For external galaxies, the cumulative covering fraction trends upward at smaller impact parameters, approaching unity at $R/R_{vir}<0.1$.

The CGM covering fraction in the Milky Way (as observed from within the Galaxy at $R\sim 0.03 R_{vir}$) is not consistent with the trend in detection rate for external galaxies. This is especially evident in the low-velocity window, which has a covering fraction of $f_{C, LVMW}=$ 20\%, well below the cumulative covering fraction of $f_{C, lit}(R<0.2 R_{vir})=$ 79\% for external galaxy sightlines. For the $|v|<300$ km/s window, the Milky Way covering fraction is higher at $f_C,MW=$ 38\% but still significantly lower than the trend for external galaxies.

It is particularly unexpected that the window that includes MS absorption has a lower covering fraction than external galaxies. Satellite galaxies the size of the Magellanic Clouds are uncommon in the halos of \lstar{} galaxies \citep{tollerud2011, liu2011, jamesivory2011}, so one might naively expect a galaxy like the Milky Way to have a covering fraction that is higher than other galaxies. The discrepancy between the Milky Way and its low-redshift neighbors only widens when we consider that three of the sightlines showing excess quasar absorption may be the result of foreground clumpiness.
Furthermore, the inconsistency persists even when we account for the shorter distance probed by the Milky Way's inside-out sightline geometry.

It is important to recognize how different observing geometries constrain CGM measurements in the Milky Way versus those made in external galaxies, and consequently how those geometries affect observed column densities.
Measurements of the Milky Way's CGM are affected by an ``inside-out" geometric observing bias that differs from the ``outside-in" geometric bias for external galaxies. Any sightline that probes the Milky Way's extended CGM must also pass through the denser and clumpier gas in the inner halo. Although the QuaStar survey is designed to circumvent this issue, subtracting absorption from foreground gas introduces some uncertainty that does not affect external galaxy measurements. Furthermore, all QuaStar sightlines are at high Galactic latitude ($|b|>$30$^\circ$) and originate from the same point within the Galaxy, while extragalactic sightlines pierce galaxy halos at a range of orientations and distances from the host galaxy.

The different observing geometries also mean that in the Milky Way, gas velocities are measured along sightlines pointing radially outward from the disk, while velocities within extragalactic systems are measured along sightlines in a direction tangential to the galaxy. In both cases, measured velocities are only a projection of the true gas velocity onto the line of sight. Therefore, even if the CGM of the Milky Way and other \lstar{} galaxies were identical, a velocity structure that changes along the radial direction would result in different observed velocities and increase the scatter of inside-out measurements by a factor of $\sim2$ \citep{zheng2020}. Such asymmetries in the kinematics of CGM gas are indeed seen in idealized simulations \citep{lochhaas2020}. If this is correct, the gas that is detected at low velocities in external galaxies might be detected at $>150$ km/s if the same gas was observed from within the galaxy \citep{zheng2015}. 

To check that our results are not being driven by an observing geometry bias, we repeated the comparison to external galaxies using a ``chord length'' parameter. The chord length of each sightline is defined as the full distance the sightline traverses within the virial radius of the host galaxy and is equal to $2\sqrt{R_{\rm vir}^2 - \rho^2}$. For external galaxy sightlines, $\rho$ is the impact parameter; for QuaStar sightlines, it is the foreground star's Galactocentric distance. Chord length therefore scales inversely with impact parameter for external galaxies. For example, in a galaxy with a virial radius of $R_{vir} = 200$ kpc, impact parameters of 50 and 150 kpc correspond to chord lengths of 387 and 265 kpc, respectively. For Milky Way sightlines with $\rho=8$ kpc, the chord length is approximately 230 kpc. If we observed the Milky Way from the outside in like an external galaxy, the same chord length would correspond to a much larger impact parameter. Even so, when we compare the Milky Way to external galaxies using chord length, we still find a significantly lower covering fraction of low-velocity \CIV in the Milky Way. The chord length thus provides an additional, conservative metric for comparing the disparate Milky Way and extragalactic datasets. However, without deprojected velocities and actual distance information for the absorbers along their sightlines, it is not possible to fully account for all biases arising from observing geometry. We emphasize that these biases may affect measured column densities in both the Milky Way and external galaxies.\\

\subsection{Understanding the \CIV Content of the Extended CGM: Is the Milky Way an Anomaly?}
\label{sec:explanation}

To interpret our results, we must consider the physical properties of \CIV-bearing gas within the broader context of a complex, multiphase CGM \citep{tumlinsonARAA2017}. Generally, \CIV~ has two possible phases: (1) along with common metal ions such as \SiIV, \NV, and \OVI, it may trace warm, collisionally ionized  ``transition temperature" or ``intermediate-ionization-state" gas at $T\sim10^{5.5}$ K \citep[e.g.][]{werk2016}; (2) \CIV (and other intermediate ions present in the CGM) could also trace low-density, cool T$\approx$ 10$^{4.5}$ K photoionized material, generally envisioned as clouds in the CGM \citep[e.g.][]{stern2016}.  In a simulated experiment similar to QuaStar, \cite{zheng2020} used mock observations of $5,000$ quasar-star pairs through the halo of a high-redshift Milky Way analog in the FOGGIE simulation \citep{peeples2019} and found that \CIV may be a less sensitive tracer of the outer CGM than \OVI and \NV. In the FOGGIE simulations of high-redshift Milky Way analogs, therefore, \CIV~appears to predominantly trace cooler material typically detected as low ionization state metal species. In contrast, the cosmological hydrodynamical simulation EAGLE, with its AGN feedback model, predicts that \CIV~manifests as both warm and cool material, making it an ideal tracer of the total baryonic content of the CGM \citep{Oppenheimer20}.   Thus, the near absence of \CIV in the QuaStar observations could mean either a lower quantity of low-density photoionized, cool clouds relative to other nearby star-forming galaxies or, alternatively, an atmosphere depleted of transitional-temperature gas, which is primarily exhibited by non-star-forming galaxies \citep{tumlinson2011}. 

With this in mind, we can speculate about what might cause the surprisingly low covering fraction we find for low-velocity gas in the Milky Way's CGM. QuaStar observations hint at a \CIV distribution that deviates notably from the norm; we find that the bulk of the column density detected in the halo beyond $d\sim14$~kpc is associated with the MS at velocities greater than 150 km/s (see \S\ref{sec:MSandLVHCs}). This is kinematically very different from sightlines through external galaxy halos, in which the bulk of the CGM gas column density tends to sit at $v<150$~km~s$^{-1}$ relative to galaxy systemic velocities \citep[e.g.,][]{tumlinson2013,werk2014,zheng2019}. These contrasting measurements do not necessarily mean that the Milky Way is missing \CIV at low velocities, however. For example, if halo gas predominantly flows in a radial direction, differing observing geometries in the Milky Way versus external galaxies (discussed in \S\ref{sec:litcomparison}) could cause halo gas velocities projected along the line of sight to be lower for external galaxies, producing a similar effect. 

Moreover, because QuaStar focuses on high-latitude ($|b|>30^\circ$) sightlines, we cannot rule out the intriguing possibility that the \CIV-traced CGM lies predominantly at lower Galactic latitudes. A toroidal or flared-disk morphology would not be probed by the QuaStar survey's inside-out high-latitude sightlines, but could be detected by outside-in sightlines through external galaxy halos with almost any orientation. Such an extended disk model is supported at least indirectly by observational and theoretical work that finds evidence for a large-scale rotational signal in the CGM \citep{hodgeskluck2016, oppenheimer2018a, martinho2019, ho2017, zabl2019}. Other work has suggested that a corotating warm gas disk in the Milky Way may be more extended in the direction perpendicular to the disk rather than the radial direction \citep[e.g.,][]{qu2019,qu2020}, so future observations using low-latitude sightlines will be important for constraining this picture of the CGM. 

We also consider the possibility that the Milky Way truly is different from most nearby star-forming galaxies. Evidence suggests that the Milky Way is a green valley galaxy, falling between the red sequence of quenched galaxies and the blue cloud of galaxies with active star formation on the color-magnitude diagram \citep{mutch2011, licquia2015, ARAAgalaxyincontext}. Star formation in Milky Way-mass galaxies declines over several Gyr on average, more slowly than lower-mass galaxies \citep{pacifici2016}. Although a global star-formation history for the Milky Way itself is difficult to determine, the Galaxy may currently be in such a period of decline \citep{snaith2014,snaith2015}. In intermediate- and high-mass galaxies at low redshift, decreasing star formation is tied to mechanisms such as gas consumption and weak tidal interactions with small satellite galaxies, like the Magellanic Clouds \citep{bell2005}. 

The transformation galaxies experience as they begin to quench is not yet fully understood, but the CGM, which plays a key role in sustaining star formation by serving as a reservoir of gas, is almost certainly impacted by the process. Indeed, the presence of \OVI in galaxy halos is confirmed to correlate with global star formation, while it is relatively absent from the halos of non-star-forming galaxies \citep{tumlinson2011}. The physical explanation for this observed dichotomy in \OVI is widely debated \citep[e.g.][]{Oppenheimer16, mcquinnwerk2018}, but if \OVI~and \CIV both trace T $\approx$ 10$^{5-6}$~K transitional-temperature gas, then the near absence of \CIV in the Milky Way's halo may be a similar reflection of its declining star-formation rate. Unfortunately, this is difficult to test directly because there are few measurements of \CIV in the CGM of quiescent galaxies available for such a comparison. Simultaneous measurements of both \OVI~and \CIV are exceedingly rare given available UV detector characteristics and the two doublets' $>$ 500 \AA~ separation in wavelength, though some such measurements have been made for a few systems \citep[e.g., ][]{werk2016, johnson2017}. 

Finally, the Local Group environment, which contains both M31 and the Milky Way, may not be representative of the typical environment of \lstar{} galaxies. Galaxy environment has been tied to the CGM properties of galaxies and may exert a divergent influence on otherwise similar galaxies \citep[e.g.][]{yoon2013, stocke14, Burchett2018}.  The SAGA survey reports that 86\% (109/127) of its Mily Way-analog satellites are star-forming, in contrast to only 29\% (4/14) of Local Group satellites of both M31 and the Milky Way in the same mass range \citep{geha2017, mao2021}. The low star formation rates of the Local Group satellites compared to other \lstar{} satellite populations may indicate different local environmental conditions shaping the Milky Way that ultimately manifest as different CGM properties.
What's more, satellites as massive as the Milky Way's SMC/LMC are not typically found around low-redshift galaxies \citep{tollerud2011, liu2011, jamesivory2011}. The Milky Way is expected to merge with the LMC in $\sim$2.4~Gyr \citep{cautun2019}, an event that will undoubtedly disrupt circumgalactic gas. After this significant interaction, the global properties of the Milky Way's CGM may better match those of nearby star-forming galaxies.

Generally, CGM covering fractions of a range of ionized and neutral gas tracers are lower in cluster environments compared to isolated galaxies \citep{yoon2013,pointon2017,zhang2019}, but assessing the nature and impact of small group environments is more difficult. In particular, data cannot yet distinguish between gas associated with an intragroup medium and the CGM of a massive halo \citep{stocke17, stocke19}. It is therefore possible that our anomalously low \CIV covering fractions and median column densities are simply indicative of a hotter intragroup medium relatively devoid of \CIV compared to other more ``typical" \lstar{} environments. The somewhat low \CIV column densities measured within 0.5 $R/R_{\rm vir}$  of M31 relative to the CGM of other nearby galaxies, evident in Figure \ref{fig:litcomparison} (yellow diamonds), may lend support to the idea that both the Milky Way and M31 reside in an environment with a depleted \CIV-traced gaseous reservoir. However, the AMIGA survey reports a \CIV covering fraction of approximately 50\% {(with a threshold column density of log N$_{\rm CIV}$ = 13.8)} within 150 kpc of M31 \citep{lehner2020}, significantly higher than what we find around the Milky Way and roughly consistent with results for other nearby \lstar{} galaxies. 

To make progress in our understanding, more observations of the low-velocity CGM will be needed. Sightlines at low Galactic latitudes in particular are important for constraining the morphology of the low-velocity CGM and testing the idea that warm gas is hidden in an extended, flared disk. HST/COS is the best available instrument for observing UV absorption lines of warm gas, and it allows us to map ions like \CIV and \SiIV. Additional measurements of these ions in the CGM of low-redshift, Milky Way-like galaxies will be key for making robust comparisons with the Milky Way. Although there are no concrete plans to bring new, highly-sensitive, high-resolution spectrographs online with wavelength coverage in the FUV, it will be necessary to measure ions that trace other gas phases to bring a complete picture of the low-velocity CGM into focus.\\

\section{Summary}
\label{sec:summary}
 
The QuaStar survey has enabled the first measurements of the Milky Way's low-velocity CGM ($|v|<150$ km/s) that are unobscured by foreground gas in the disk. We measured \CIV column densities along sightlines to 30 halo stars (d $\sim$ 8 kpc) paired with quasars at small angular separations ($<2.8^\circ$), evenly sampling the sky at Galactic latitudes $|b|>30^\circ$ (\S\ref{sec:observations}; Figure \ref{fig:experimentaldesign}). Applying a spectral differencing technique, we isolated Milky Way CGM absorption in each quasar spectrum by subtracting the foreground gas absorption measured in its paired stellar spectrum to obtain a ``difference measurement" \diffLVCGM (\S\ref{sec:analysis}; Figure \ref{fig:difference-spectra}). Our main results are as follows:

\begin{enumerate}
    \item We detect low-velocity ($|v|<150$ km/s) \CIV in the Milky Way's CGM for 6/30 sightline pairs, place an upper limit of \diffLVCGM$<13.39$ cm$^{-2}$ on the median column density, and report a covering fraction of $f_{\rm CIV, LVCGM}$(logN$>$13.65) = 20\% [6/30]. Within a wider velocity range that includes the Magellanic System ($|v|<300$ km/s), we detect \CIV for 10/30 sightline pairs, place an upper limit of logN$<13.56$ cm$^{-2}$ on the median column density and report a covering fraction of $f_{\rm CIV, CGM}$(logN$>$13.65) = 38\% [9/24]. A total of 96\% of the additional absorption detected in this wider velocity window likely originates in the Magellanic System. Accounting for the existence of foreground substructure with angular scales below our sightline separations further reduces these two covering fractions to 10\% and 25\%, respectively.  (\S\ref{sec:analysis}, \S\ref{sec:results}, and \S\ref{sec:MSandLVHCs}; Figures \ref{fig:difference-spectra} and \ref{fig:diffplots-combined})
    
    \item The covering fraction of low-velocity \CIV in the Milky Way's CGM is markedly lower than observations of other \lstar{} galaxies. Although it is possible that the Milky Way represents an anomaly, the differing geometric viewing effects of the Milky Way and external galaxies allow for the possibility that the warm, low-velocity CGM of \lstar{} galaxies exhibits a flared-disk morphology. QuaStar's ``inside-out'' observations of the Milky Way at high latitudes would not probe such a structure, while ``outside-in'' sightlines with various orientations would be more likely to detect it in external galaxies. Alternatively, it is possible that the Milky Way's CGM is uniquely affected by environment, or has begun the process of quenching and thus appears different from typical star-forming galaxies. 
    (\S\ref{sec:discussion}; Figures \ref{fig:litcomparison} and \ref{fig:coveringfractions})\\
    
\end{enumerate}

\section{Acknowledgements}

Support for this work was provided by NASA through program GO-15656. J.K.W. and H.V.B. acknowledge additional support from the Alfred P. Sloan Foundation, Grant No: FG-2018-10555, and the National Science Foundation, grants PHY-1748958 and AST-1812531. This work has made use of observations taken with the NASA/ESA Hubble Space Telescope, and obtained from the Hubble Legacy Archive, which is a collaboration between the Space Telescope Science Institute (STScI/NASA), the Space Telescope European Coordinating Facility (ST-ECF/ESAC/ESA), and the Canadian Astronomy Data Centre (CADC/NRC/CSA). We thank the anonymous referee for thoughtful and constructive feedback that has improved this work.

H.V.B. and J.K.W. recognize the unceded traditional lands of the Duwamish and Puget Sound Salish Tribes, on which we are grateful to live and work. In addition, the authors acknowledge Guido Münch, who passed away at 98 as this manuscript was being drafted and whose pioneering work on \NaI- and \CaII-bearing gas in the Milky Way has made our research possible.

This work has made use of data from the European Space Agency (ESA) mission Gaia (\url{https://www.cosmos.esa.int/gaia}), processed by the Gaia Data Processing and Analysis Consortium (DPAC, \url{https://www.cosmos.esa.int/web/gaia/dpac/consortium}). Funding for the DPAC has been provided by national institutions, in particular the institutions participating in the Gaia Multilateral Agreement.

The national facility capability for SkyMapper has been funded through ARC LIEF grant LE130100104 from the Australian Research Council, awarded to the University of Sydney, the Australian National University, Swinburne University of Technology, the University of Queensland, the University of Western Australia, the University of Melbourne, Curtin University of Technology, Monash University, and the Australian Astronomical Observatory. SkyMapper is owned and operated by the Australian National University's Research School of Astronomy and Astrophysics. The survey data were processed and provided by the SkyMapper Team at ANU. The SkyMapper node of the All-Sky Virtual Observatory (ASVO) is hosted at the National Computational Infrastructure (NCI). Development of and support for the SkyMapper node of the ASVO have been funded in part by Astronomy Australia Limited (AAL) and the Australian Government through the Commonwealth's Education Investment Fund (EIF) and National Collaborative Research Infrastructure Strategy (NCRIS), particularly the National eResearch Collaboration Tools and Resources (NeCTAR) and the Australian National Data Service Projects (ANDS).

\facility{HST: COS}


\bibliographystyle{apj}
\bibliography{./QuaStar}
\clearpage


\input{table}
\clearpage

\end{document}

%% file: table.tex
\LongTables
\renewcommand{\arraystretch}{1.7}
\begin{deluxetable*}{l l l l r r l c l l r}
\tabletypesize{\scriptsize}
\tablecaption{BHB-QSO Sightline Pairs}
\tablehead{
\colhead{ID  \vspace{0pt}} &
\colhead{Name} &
\colhead{RA} &
\colhead{Dec} &
\colhead{$l$} &
\colhead{$b$} &
\colhead{$\Delta \theta$} &
\colhead{$d_{star}$} &
\colhead{$\rm logN_{CIV}$} &
\colhead{$\rm logN_{LVCGM}$} &
\colhead{\vspace{-2pt}} \\
\colhead{\vspace{0pt}} &
\colhead{} &
\colhead{[h:m:s]} & 
\colhead{[d:m:s]} & 
\colhead{[$^{\circ}$]} & 
\colhead{[$^{\circ}$]} & 
\colhead{[$^{\circ}$]} & 
\colhead{[kpc]} & 
\colhead{[\percmsq]} & 
\colhead{[\percmsq]} &
\colhead{} \\
\colhead{(1)} &
\colhead{(2)} &
\colhead{(3)} & 
\colhead{(4)} &
\colhead{(5)} &
\colhead{(6)} & 
\colhead{(7)} & 
\colhead{(8)} &
\colhead{(9)} & 
\colhead{(10)} &
\colhead{(11)}
}
\startdata
1S  & J1339+3136               & 13:39:05.82  & +31:36:30.70    & 59.12  & 78.70     & \multirow{2}{*}{1.77} & 9.6   & $14.19^{+0.05}_{-0.06}$   & \multirow{2}{*}{$<$13.34}  & \multirow{2}{*}{$\varnothing$}    \\
1Q  & SDSSJ133053.27+311930.5  & 13:30:53.28  & +31:19:30.72    & 61.28  & 80.43     &                       & -     & $14.17^{+0.05}_{-0.06}$   &                            &                 \\
\hline
2S & J1344+1842                & 13:44:04.41 & +18:42:59.03    & 0.89   & 75.26     & \multirow{2}{*}{0.31} & 10.4  & $14.37^{+0.03}_{-0.03}$   & \multirow{2}{*}{$<$13.47}  & \multirow{2}{*}{$\varnothing$}    \\
2Q & SDSSJ134246.89+184443.6   & 13:42:46.80 & +18:44:43.80    & 0.24   & 75.53     &                       & -     & $14.37^{+0.06}_{-0.07}$   &                            &       \\
\hline 
3S  & J1406+2441               & 14:06:38.54  & +24:41:41.89    & 28.96  & 72.93     & \multirow{2}{*}{1.86} & 9.0   & $14.26^{+0.05}_{-0.05}$   & \multirow{2}{*}{$<$13.32}  & \multirow{2}{*}{$\varnothing$}      \\
3Q  & SDSS-J141038.39+230447.1 & 14:10:38.40  & +23:04:47.18    & 24.57  & 71.64     &                       & -     & $14.25^{+0.04}_{-0.04}$   &                            &     \\
\hline    
4S  & J1131+3112               & 11:31:39.17  & +31:12:47.98    & 194.96 & 72.14     & \multirow{2}{*}{0.11} & 8.2   & $14.11^{+0.06}_{-0.07}$   & \multirow{2}{*}{$<$13.26}  & \multirow{2}{*}{$\varnothing$}       \\
4Q  & TON580                   & 11:31:09.50  & +31:14:05.00    & 194.94 & 72.03     &                       & -     & $14.12^{+0.02}_{-0.03}$   &                            &       \\
\hline   
5S  & J1113+2736               & 11:13:47.22  & +27:36:48.51    & 205.99 & 68.35     & \multirow{2}{*}{1.39} & 11.4  & $14.11^{+0.06}_{-0.07}$   & \multirow{2}{*}{+$13.73^{+0.23}_{-0.52}$}    & \multirow{2}{*}{\QSOexcessdot} \\
5Q  & SDSSJ111754.31+263416.6  & 11:17:54.24  & +26:34:16.68    & 209.10 & 69.16     &                       & -     & $14.26^{+0.07}_{-0.08}$   &                            &       \\
\hline   
6S  & J1212+0621               & 12:12:21.95  & +06:21:24.30    & 277.09 & 67.25     & \multirow{2}{*}{1.76} & 7.3   & $13.96^{+0.06}_{-0.07}$   & \multirow{2}{*}{+$14.07^{+0.07}_{-0.08}$}    & \multirow{2}{*}{\QSOexcessdot} \\
6Q  & PG1216+069               & 12:19:20.93  & +06:38:38.52    & 281.07 & 68.14     &                       & -     & $14.32^{+0.03}_{-0.04}$   &                            &       \\
\hline    
7S  & J1118+4029               & 11:18:21.44  & +40:29:07.88    & 172.15 & 66.62     & \multirow{2}{*}{1.94} & 7.9   & $14.10^{+0.07}_{-0.08}$   & \multirow{2}{*}{$<$13.34}  & \multirow{2}{*}{$\varnothing$}      \\
7Q  & PG1121+422               & 11:24:39.18  & +42:01:45.02    & 167.26 & 66.86     &                       & -     & $14.13^{+0.04}_{-0.05}$   &                            &       \\
\hline    
8S  & J1339+0511               & 13:39:03.97  & +05:11:57.35    & 332.40 & 65.31     & \multirow{2}{*}{0.77} & 8.2   & $14.22^{+0.06}_{-0.07}$   & \multirow{2}{*}{$<$13.37}  & \multirow{2}{*}{$\varnothing$}      \\
8Q  & SDSSJ134206.56+050523.8  & 13:42:06.48  & +05:05:24.00    & 333.89 & 64.87     &                       & -     & $14.21^{+0.04}_{-0.04}$   &                            &       \\
\hline    
9S  & J1449+3459               & 14:49:20.37  & +34:59:40.18    & 57.61  & 63.67     & \multirow{2}{*}{1.00} & 7.1   & $14.19^{+0.05}_{-0.06}$   & \multirow{2}{*}{$<$13.64}  & \multirow{2}{*}{$\varnothing$}      \\
9Q  & SDSSJ144511.28+342825.4  & 14:45:11.28  & +34:28:25.45    & 56.74  & 64.59     &                       & -     & $14.14^{+0.13}_{-0.19}$   &                            &          \\
\hline    
10S  & J1400+0535               & 14:00:35.61  & +05:35:17.21    & 343.40 & 62.89     & \multirow{2}{*}{1.27} & 7.7   & $13.96^{+0.09}_{-0.11}$   & \multirow{2}{*}{$<$13.36}  & \multirow{2}{*}{$\varnothing$}       \\
10Q  & SDSS-J135726.27+043541.4 & 13:57:26.27  & +04:35:41.40    & 340.77 & 62.51     &                       & -     & $13.99^{+0.07}_{-0.09}$   &                            &         \\
\hline    
11S & J1235-0108               & 12:35:00.48  & -01:08:54.80    & 294.32 & 61.45     & \multirow{2}{*}{0.79} & 7.1   & $14.09^{+0.06}_{-0.07}$   & \multirow{2}{*}{+$14.11^{+0.12}_{-0.17}$}    & \multirow{2}{*}{\QSOexcessdot} \\
11Q & SDSSJ123304.05-003134.1  & 12:33:04.05  & -00:31:34.17    & 293.11 & 61.99     &                       & -     & $14.40^{+0.06}_{-0.07}$   &                            &       \\
\hline 
12S & PG0955+291               & 09:58:15.14  & +28:52:33.12    & 199.88 & 51.94     & \multirow{2}{*}{1.27} & 5.5   & $14.17^{+0.02}_{-0.03}$   & \multirow{2}{*}{$<$13.09}  & \multirow{2}{*}{$\varnothing$}      \\
12Q & PG1001+291               & 10:04:02.59  & +28:55:35.18    & 200.08 & 53.21     &                       & -     & $14.13^{+0.03}_{-0.03}$   &                            &       \\
\hline     
13S & J1057-0108               & 10:57:52.64  & -01:08:54.95    & 254.32 & 50.68     & \multirow{2}{*}{1.53} & 9.1   & $<13.28$                  & \multirow{2}{*}{$<$13.41}  & \multirow{2}{*}{$\varnothing$}      \\
13Q & PG1049-005               & 10:51:51.44  & -00:51:17.73    & 252.28 & 49.88     &                       & -     & $13.47^{+0.22}_{-0.45}$   &                            &       \\
\hline    
14S & J1511+0452               & 15:11:38.17  & +04:52:54.54    & 5.56   & 49.89     & \multirow{2}{*}{1.84} & 10.8  & $14.41^{+0.05}_{-0.05}$   & \multirow{2}{*}{$-13.88^{+0.24}_{-0.15}$}    & \multirow{2}{*}{\BHBexcessdot}  \\
14Q & MRK1392                  & 15:05:56.55  & +03:42:26.21    & 2.75   & 50.26     &                       & -     & $14.26^{+0.02}_{-0.03}$   &                            &       \\
\hline    
15S & J1019-0051               & 10:19:35.54  & +00:51:12.57    & 244.15 & 44.01     & \multirow{2}{*}{2.49} & 7.5   & $13.88^{+0.10}_{-0.13}$   & \multirow{2}{*}{$<$13.56}  & \multirow{2}{*}{$\varnothing$}       \\
15Q & SDSSJ102218.99+013218.8  & 10:22:18.99  & +01:32:18.82    & 242.16 & 46.07     &                       & -     & $13.90^{+0.15}_{-0.23}$   &                            &       \\
\hline    
16S & J0841+4529               & 08:41:21.54  & +45:29:12.42    & 174.86 & 37.81     & \multirow{2}{*}{2.74} & 10.4  & $13.85^{+0.10}_{-0.13}$   & \multirow{2}{*}{+$13.93^{+0.09}_{-0.12}$}    & \multirow{2}{*}{\QSOexcessdot} \\
16Q & Q0850+440                & 08:53:34.24  & +43:49:02.28    & 177.08 & 39.94     &                       & -     & $14.20^{+0.02}_{-0.02}$   &                            &       \\
\hline    
17S & J0841+2340               & 08:41:40.74  & +23:40:03.36    & 201.51 & 34.07     & \multirow{2}{*}{1.92} & 8.5   & $13.61^{+0.18}_{-0.33}$   & \multirow{2}{*}{+$13.94^{+0.15}_{-0.22}$}    & \multirow{2}{*}{\QSOexcessdot} \\
17Q & PG0832+251               & 08:35:35.80  & +24:59:41.00    & 199.49 & 33.15     &                       & -     & $14.11^{+0.09}_{-0.11}$   &                            &       \\
\hline
18S & J0456-2447              & 04:56:15.22  & -24:47:36.77    & 225.26 & -35.48    & \multirow{2}{*}{2.81} & 13.5  & $13.79^{+0.12}_{-0.17}$   & \multirow{2}{*}{$<$13.31}  & \multirow{2}{*}{$\varnothing$}    \\
18Q & QSOJ0456-2159           & 04:56:08.93  & -21:59:09.30    & 221.98 & -34.65    &                       & -     & $13.84^{+0.07}_{-0.08}$   &                            &       \\
\hline    
19S & J2129-1112               & 21:29:17.55  & -11:12:03.19    & 41.30  & -40.05    & \multirow{2}{*}{1.08} & 11.1  & $14.37^{+0.05}_{-0.05}$   & \multirow{2}{*}{$-13.76^{+0.27}_{-0.16}$}    & \multirow{2}{*}{\BHBexcessdot}  \\
19Q & PHL1598                  & 21:31:35.20  & -12:07:04.50    & 40.54  & -40.96    &                       & -     & $14.24^{+0.02}_{-0.02}$   &                            &       \\
\hline 
20S & J0158-7527               & 01:58:15.71  & -75:27:25.37    & 297.45 & -40.95    & \multirow{2}{*}{0.91} & 8.8   & $14.20^{+0.06}_{-0.07}$   & \multirow{2}{*}{$-13.70^{+0.80}_{-0.26}$}    & \multirow{2}{*}{{\tiny MS} \BHBexcessdot}     \\
20Q & HB89-0202-765            & 02:02:13.70  & -76:20:03.10    & 297.55 & -40.05    &                       & -     & $14.03^{+0.12}_{-0.16}$   &                            &       \\
21S & J0420-5233               & 04:20:08.33  & -52:33:47.75    & 261.11 & -43.93    & \multirow{2}{*}{2.56} & 8.5   & $14.31^{+0.05}_{-0.05}$   & \multirow{2}{*}{$<$13.43}  & \multirow{2}{*}{$\varnothing$}     \\
21Q & HE0435-5304              & 04:36:50.80  & -52:58:49.00    & 261.02 & -41.37    &                       & -     & $14.29^{+0.05}_{-0.06}$   &                            &       \\
\hline  
22S & J2326-7108               & 23:26:01.06  & -71:08:39.21    & 312.41 & -44.38    & \multirow{2}{*}{0.78} & 10.6  & $14.22^{+0.05}_{-0.06}$   & \multirow{2}{*}{$<$13.37}  & \multirow{2}{*}{$\varnothing$}      \\
22Q & RXS-J23218-7026          & 23:21:51.10  & -70:26:44.00    & 313.29 & -44.84    &                       & -     & $14.24^{+0.03}_{-0.03}$   &                            &       \\
\hline  
23S & J2156-4403               & 21:56:38.48  & -44:03:02.93    & 355.37 & -51.21    & \multirow{2}{*}{0.37} & 8.5   & $14.21^{+0.06}_{-0.06}$   & \multirow{2}{*}{$<$13.32}  & \multirow{2}{*}{$\varnothing$}      \\
23Q & RXJ2154.1-4414           & 21:54:51.06  & -44:14:06.00    & 355.18 & -50.86    &                       & -     & $14.18^{+0.03}_{-0.03}$   &                            &       \\
\hline  
24S & J0228-0132               & 02:28:23.16  & -01:32:41.41    & 169.27 & -55.44    & \multirow{2}{*}{1.88} & 8.2   & $14.18^{+0.06}_{-0.07}$   & \multirow{2}{*}{$<$13.50}  & \multirow{2}{*}{$\varnothing$}     \\
24Q & GALEX-J022614.4+001530   & 02:26:14.46  & +00:15:30.01    & 166.57 & -54.38    &                       & -     & $14.10^{+0.08}_{-0.10}$   &                            &       \\
\hline   
25S & J0243-0636               & 02:43:43.26  & -06:36:59.69    & 180.49 & -56.37    & \multirow{2}{*}{1.39} & 8.8   & $14.00^{+0.09}_{-0.12}$   & \multirow{2}{*}{$<$13.56}  & \multirow{2}{*}{$\varnothing$}      \\
25Q & SDSSJ024250.85-075914.2  & 02:42:50.87  & -07:59:14.28    & 182.15 & -57.42    &                       & -     & $14.00^{+0.11}_{-0.16}$   &                            &       \\
\hline   
26S & J2345-0157               & 23:45:29.55  & -01:57:28.50    & 88.03  & -60.28    & \multirow{2}{*}{0.97} & 8.5   & $14.04^{+0.07}_{-0.08}$   & \multirow{2}{*}{$<$13.55}  & \multirow{2}{*}{{\tiny MS} $\varnothing$}    \\
26Q & SDSSJ234500.43-005936.0  & 23:45:00.43  & -00:59:36.06    & 88.79  & -59.39    &                       & -     & $14.07^{+0.11}_{-0.16}$   &                            &       \\
\hline     
27S & J2255-1727               & 22:55:27.05  & -17:27:10.53    & 46.73  & -61.56    & \multirow{2}{*}{0.35} & 7.1   & $14.19^{+0.06}_{-0.07}$   & \multirow{2}{*}{$<$13.43}  & \multirow{2}{*}{{\tiny MS} $\varnothing$}    \\
27Q & MR2251-178               & 22:54:05.88  & -17:34:55.30    & 46.20  & -61.33    &                       & -     & $14.19^{+0.02}_{-0.02}$   &                            &       \\
\hline     
28S & J0235-1757               & 02:35:31.78  & -17:57:34.98    & 197.48 & -64.34    & \multirow{2}{*}{1.49} & 10.9  & $14.08^{+0.08}_{-0.09}$   & \multirow{2}{*}{$<$13.33}  & \multirow{2}{*}{$\varnothing$}      \\
28Q & HE0238-1904              & 02:40:32.50  & -18:51:51.00    & 200.48 & -63.63    &                       & -     & $14.00^{+0.03}_{-0.03}$   &                            &       \\
\hline     
29S & J0004-0845               & 00:04:41.65  & -08:45:06.77    & 89.73  & -68.56    & \multirow{2}{*}{2.50} & 9.4   & $13.97^{+0.06}_{-0.08}$   & \multirow{2}{*}{$<$13.46}  & \multirow{2}{*}{{\tiny MS} $\varnothing$}    \\
29Q & NEWQZ026                 & 00:12:24.01  & -10:22:26.40    & 92.32  & -70.89    &                       & -     & $13.93^{+0.13}_{-0.18}$   &                            &       \\
\hline     
30S & J0053-3606               & 00:53:39.63  & -36:06:49.70    & 300.06 & -81.00    & \multirow{2}{*}{1.00} & 6.5   & $14.09^{+0.07}_{-0.08}$   & \multirow{2}{*}{+$13.98^{+0.10}_{-0.12}$}    & \multirow{2}{*}{{\tiny MS} \QSOexcessdot}   \\
30Q & HE0056-3622              & 00:58:37.39  & -36:06:05.03    & 293.72 & -80.90    &                       & -     & $14.34^{+0.02}_{-0.02}$   &                            &       
\enddata
\tablecomments{Properties of star-quasar sightline pairs. (1) Sightline pair ID used in this work, followed by ``S'' for star and ``Q'' for quasar. (2) Catalog name of the star or quasar. (3 - 6) Right ascension/declination and Galactic longitude/latitude of the source. (7) Angular separation on the sky between the star and quasar sightlines. (8) Distance to the star. No distance is reported for quasars. (9) Column densities for the \CIV ($\lambda\lambda$ 1548, 1550 \AA) doublet within $\pm 150$ $\kms$  of the Local Standard of Rest (LSR). (10) The difference in the measured \CIV column densities for each close sightline pair. (11) Symbols corresponding to classification of low-velocity column density difference measurement: $\varnothing$\hspace{-1.pt} non-detection, \QSOexcessdot\hspace{-1.pt} excess QSO absorption, or \BHBexcessdot\hspace{-1.5pt} excess stellar absorption. Sightlines in the direction of the Magellanic Stream are labeled ``MS''.
}
\label{table:properties}
\end{deluxetable*}